 \documentclass[sigconf]{acmart} 

\renewcommand\footnotetextcopyrightpermission[1]{} 
\AtBeginDocument{%
  }

\setcopyright{acmlicensed}
\copyrightyear{2018}
\acmYear{2018}
\acmDOI{XXXXXXX.XXXXXXX}
\acmConference[Conference acronym 'XX]{Make sure to enter the correct
  conference title from your rights confirmation emai}{June 03--05,
  2018}{Woodstock, NY}
\acmISBN{978-1-4503-XXXX-X/18/06}

\definecolor{cBLUE}{HTML}{3282B8}
\definecolor{cGREEN}{HTML}{60A561}
\definecolor{cORANGE}{HTML}{FA824C}
\definecolor{cYELLOW}{HTML}{f0C808}
\definecolor{cLightGrey}{HTML}{CECECE}
\definecolor{cRED}{HTML}{ED1B23}

\definecolor{cNatureORANGE}{HTML}{EAAA60}
\definecolor{cNatureRED}{HTML}{E68B81}
\definecolor{cNaturePURPLE}{HTML}{B7B2D0}
\definecolor{cNatureBLUE}{HTML}{7DA6C6}
\definecolor{cNatureGREEN}{HTML}{84C3B7}

\begin{document}

\title[Design Opportunities of AI Remote Monitoring for Mental Health Sequelae in Youth Concussion Patients]{More Modality, More AI: Exploring Design Opportunities of AI-Based Multi-modal Remote Monitoring Technologies for Early Detection of Mental Health Sequelae in Youth Concussion Patients}


\author{Bingsheng Yao}
\authornote{Both authors contributed equally to this research.}
\author{Menglin Zhao}
\authornotemark[1]
\affiliation{
  \institution{Northeastern University}
  \city{}
  \state{}
  \country{}
}


\author{Weidan Cao}
\affiliation{Ohio State University
  \institution{}
  \city{}
  \state{}
  \country{}}
\email{}

\author{Changchang Yin}
\affiliation{Ohio State University
  \institution{}
  \city{}
  \state{}
  \country{}}
\email{}

\author{Stephen Intille}
\affiliation{Northeastern University
  \institution{}
  \city{}
  \state{}
  \country{}}
\email{}

\author{Xuhai "Orson" Xu}
\affiliation{Columbia University
  \institution{}
  \city{}
  \state{}
  \country{}}
\email{}

\author{Ping Zhang}
\affiliation{Ohio State University
  \institution{}
  \city{}
  \state{}
  \country{}}
\email{}

\author{Jingzhen Yang}
\affiliation{Nationwide Children’s Hospital
  \institution{}
  \city{}
  \state{}
  \country{}
  \institution{}
  \city{}
  \state{}
  \country{}}
\email{}

\author{Yuling Sun}
\affiliation{University of Michigan
  \institution{}
  \city{}
  \state{}
  \country{}}
\email{}

\author{Dakuo Wang}
\authornote{Corresponding author: d.wang@northeastern.edu}
\affiliation{Northeastern University
  \institution{}
  \city{}
  \state{}
  \country{}}
\email{}

\renewcommand{\shortauthors}{Yao and Zhao et al.}

\begin{abstract}
Anxiety, depression, and suicidality are common mental health sequelae following concussion in youth patients, often exacerbating concussion symptoms and prolonging recovery. 
Despite the critical need for early detection of these mental health symptoms, clinicians often face challenges in accurately collecting patients' mental health data and making clinical decision-making in a timely manner.
Today's remote patient monitoring (RPM) technologies offer opportunities to objectively monitor patients’ activities, but they were not specifically designed for youth concussion patients; moreover, the large amount of data collected by RPM technologies may also impose significant workloads on clinicians to keep up with and use the data. 
To address these gaps, we employed a three-stage study consisting of a formative study, interface design, and design evaluation.
We first conducted a formative study through semi-structured interviews with six highly professional concussion clinicians and identified clinicians' key challenges in remotely collecting patient information and accessing patient treatment compliance.
Subsequently, we proposed preliminary clinician-facing interface designs with the integration of AI-based RPM technologies (\textbf{AI-RPM}), followed by design evaluation sessions with highly professional concussion clinicians. 
Clinicians underscored the value of integrating multi-modal AI-RPM technologies to support their decision-making while emphasizing the importance of customizable interfaces through collaborative design and multiple responsible design considerations.
\end{abstract}



\begin{teaserfigure}
  \vspace{-1em}
  \centering
  \includegraphics[draft=false,width=.93\textwidth]{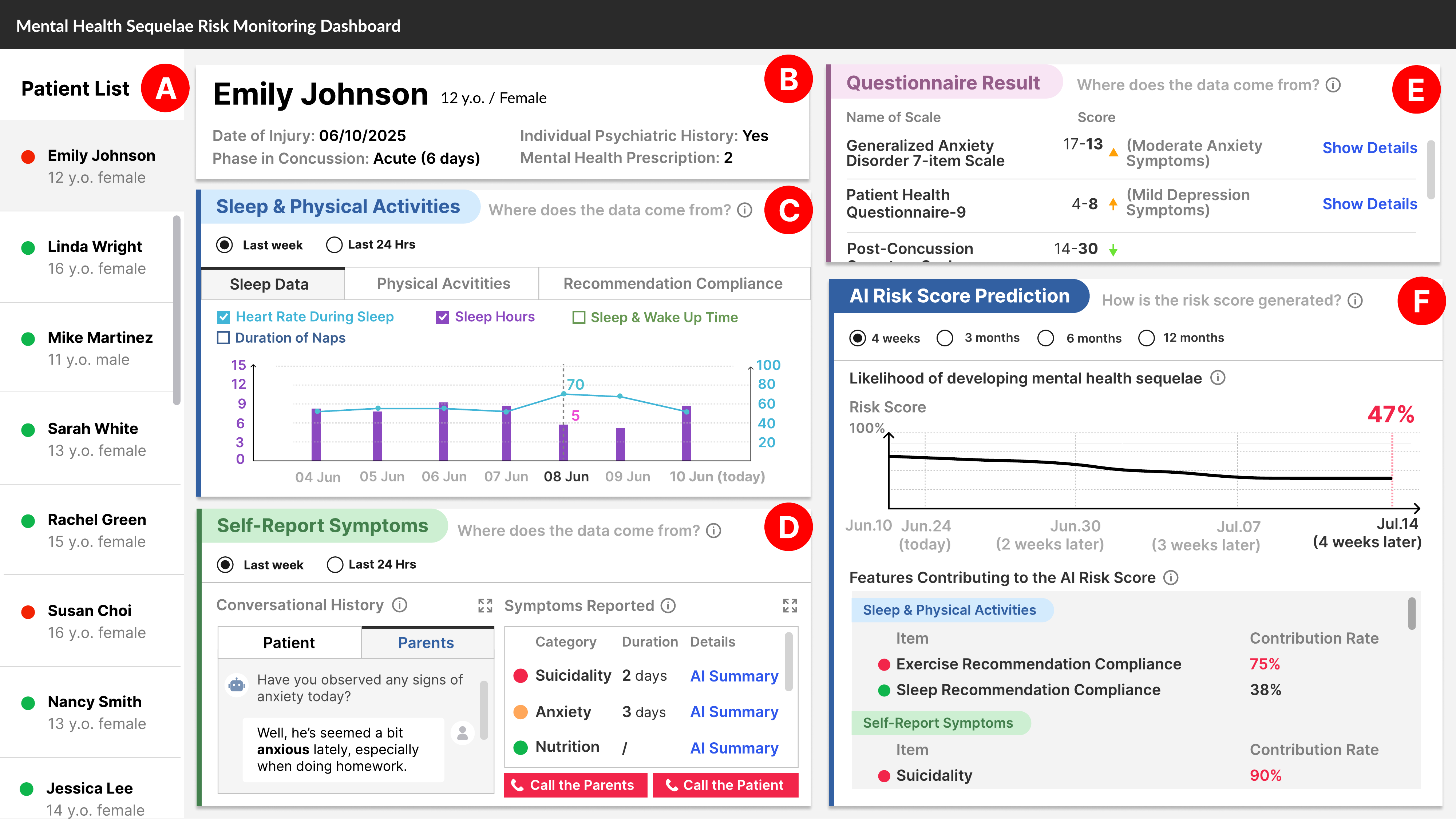}
  \vspace{-0.5em}
  \caption{Refined AI-RPM system designed to support concussion clinicians in mental health monitoring: (A) patient list, (B) patient basic information, (C) sleep and physical activities, (D) self-report symptoms, (E) questionnaire result, and (F) AI risk score prediction for mental health sequelae.
}
  \label{fig:refined_system_design}
\end{teaserfigure}

\received{20 February 2007}
\received[revised]{12 March 2009}
\received[accepted]{5 June 2009}

\maketitle

\section{Introduction}
Concussions are a form of traumatic brain injury (TBI) caused by a sudden hit or jolt of the head.
In 2023, a total of $2.3$ million youth aged $\leq17$ years received a diagnosis of concussion, making it one of the most common injuries among youth patients aged 11 to 17~\cite{yang2021association,miller2022salivary,rivara2014sports,bryan2016sports}.
Concussions can often lead to a high likelihood of developing mental health disorders (i.e.,~\textbf{mental health sequelae}) among youth patients, such as depression, anxiety, post-traumatic stress disorder (PTSD), and suicidality~\cite{fralick2019association,ceniti2022psychological,gornall2021mental}.
Mental health sequelae can, in turn, exacerbate the severity of concussion symptoms, including insomnia, concentration difficulties, and headaches, which may result in prolonged concussion recovery~\cite{gornall2021mental,iverson2003examination,ceniti2022psychological} and hospitalization~\cite{ledoux2022risk}.
Researchers have emphasized the importance of early detection of mental health symptoms in youth concussion patients~\cite{miller2021association, brooks2019predicting, gornall2021mental, brent2017psychiatric}
The early symptoms of mental health sequelae (e.g., disturbed sleep patterns, fatigue, and emotional changes ~\cite{brent2017psychiatric, silverberg2023management}) are often difficult to observe as these symptoms typically occur outside the clinics.
Concussion patients' infrequent clinical visits (e.g., every two weeks) further exacerbate the difficulty for clinicians to observe such symptoms and detect mental health sequelae in a timely manner~\cite{silverberg2023management,rose2015diagnosis,mcleod2017rest}.

The standard of care (SoC) approaches of concussion clinicians include: (1) asking patients to recall their at-home behaviors during clinical visits~\cite{silverberg2020management,brent2017psychiatric}, and (2) instructing patients or parents to complete self-evaluation questionnaires at home, such as Generalized Anxiety Disorder-7 (GAD-7) and Patient Health Questionnaire-9 (PHQ-9)~\cite{carson2021using, silverberg2020management}.
However, these approaches are not without limitations:
it could be challenging for concussion patients to comprehensively and accurately recall their previous behaviors in a clinic visit~\cite{frissa2016challenges,van2016accuracy}; 
patients may be able to complete questionnaires on a regular basis at home, but concussion clinicians can only access the questionnaire results when patients return to the clinics in person, which still delays clinicians' decision-making.

Previous work has proposed different Remote Patient Monitoring (RPM) technologies to collect patients' at-home information in various scenarios~\cite{chung2016boundary,olen2017implantable,barnes2024clinician,van2023clinician,toresdahl2021systematic,tirosh2024smartphone}.
For example, mobile health (mHealth) applications have been proposed in which patients can report their physiological measurements to care provider teams in real time. 
Wearable devices, such as the ActiGraph~\cite{yang2020bidirectional} and Philips Actiwatch~\cite{donahue2024feasibility}, have been explored to measure the daily steps and sleep patterns of adult concussion patients.
Despite their relative success in capturing patients' activity data, these traditional RPM approaches are not without limitations: mHealth applications could lead to cognitive burdens for patients with low digital literacy and brain injuries~\cite{giebel2024problems,purdy2023exploring}; extensive amount of time and effort is required to process the massive amount of multi-modal data collected by wearable devices~\cite{ginsburg2024key,baig2017systematic}.
These limitations may reduce patient compliance and lead to clinicians delaying collected data access,  which further affects timely diagnosis and intervention.

Recent exploration in the HCI community demonstrates the promising potential of leveraging innovative AI technologies for RPM (\textbf{AI-RPM}) to address the aforementioned limitations of traditional RPM approaches. 
For instance, conversational agents (CAs) powered by large language models (LLMs) enable asynchronous patient-provider communication to verbally collect patients' health conditions through regular health check-in conversations~\cite{yang2024talk2care}. 
LLMs have also been explored to organize and summarize key information from massive RPM data on clinician-facing dashboards~\cite{wu2024cardioai}.
Meanwhile, another line of research explored using AI prediction algorithms with RPM data to assist clinicians' decision-making in their clinical workflow~\cite{yin2024sepsislab}.

Despite the promising opportunities of AI-RPM systems, they could lead to additional difficulties and burdens if not appropriately designed.
Recent studies discovered that AI prediction models without faithful and reliable explanations cannot provide helpful information~\cite{antoniadi2021current} or actionable insights to clinicians~\cite{rane2023explainable, zhang2024rethinking}, lose clinician trust~\cite{zhang2020effect, chanda2024dermatologist}, and even result in human-AI competition~\cite{zhang2024rethinking}.
Moreover, in high-stakes, high-risk, and uncertain clinical scenarios, clinicians follow strict clinical guidelines and specifications in their daily workflow. 
AI systems misaligned with clinical specifications may be unusable and even cause potential harm to stakeholders~\cite{romero2020lesson}.

In contrast, human-centered AI (HCAI) design guidelines~\cite{shneiderman2022human, wang2023human} underscore the importance of engaging stakeholders in the system design process and understanding stakeholders' workflow, in particular, the needs and challenges they encounter.
By following HCAI design guidelines, AI systems can be seamlessly integrated into stakeholders' workflow where stakeholders can confidently understand, trust, and use AI-powered systems in their current workflow. 
A number of studies have demonstrated the effectiveness of human-centered AI systems in medical and healthcare scenarios to facilitate the needs of diverse stakeholders, including older adults~\cite{hao2024advancing}, cancer patients~\cite{wu2024cardioai}, and clinicians~\cite{wang2023human,fogliato2022goes,yang2019unremarkable,zhang2024rethinking}.
Nevertheless, how to employ HCAI guidelines in the design of AI-RPM systems has not been studied yet, and little is known about the challenges concussion clinicians face in remote monitoring youth patients and how AI-RPM technologies could be of practical benefit.

In this work, our objective is to investigate the current workflow of concussion clinicians with respect to the challenges in remote monitoring and decision-making of young concussion patients.
In particular, we focus on how clinicians collect patients' mental health-related information, what decisions they need to make, and what the difficulties are in their current workflow.
In addition, we want to explore the design opportunities of AI-RPM technologies in supporting clinicians' needs for remote monitoring and decision-making during their practice.
To answer these questions, we conducted a formative study of semi-structured interviews with six highly professional concussion clinicians and derived a suite of design considerations of AI-RPM technologies that could support clinicians' needs and clinical decision-making.
We then came up with preliminary system designs based on design considerations.
Subsequently, we engaged the same group of clinicians in an evaluation study to further collect feedback and insights on the design.
Finally, we consolidate the feedback and revise the design to come up with a complete clinician-facing system design as the final artifact. 

This paper makes the following key contributions:
(1) we identified the challenges and needs that concussion clinicians encounter during their clinical practice with youth concussion patients, particularly with respect to patient mental health sequelae; 
(2) we derive a suite of design considerations and visualization feedback to empower advanced AI-RPM technologies to support remote patient monitoring and clinicians' decision-making; and 
(3) we deliver a user-centered system design of a concussion clinician-oriented dashboard system supported by multiple AI-RPM technologies, where the preliminary system design and refinements are grounded in users' needs and feedback.

\section{Related Work}
Based on the prior research, we first identified the challenges related to mental health sequelae after a concussion. 
Then, we provided a brief overview of current AI-RPM technologies in clinical settings, as well as AI-based system design for clinical decision support.

\subsection{Challenges in Post-Concussion Mental Health Sequelae  }
Mental health sequelae, such as anxiety, depression, PTSD, and suicidality, are common among youth concussion patients (ages 11–17)~\cite{stein2019risk, gornall2021mental, ledoux2022risk}. 
Yang et al.~\cite{yang2015post} studied how concussions impacted the mental health of 71 collegiate athletes and found that 14 reported symptoms of depression, 24 experienced symptoms of anxiety, and 10 reported both depression and anxiety. 
Another study showed that youth concussion patients tend to have twice the risk of suicide compared to their peers who have never had a concussion~\cite{fralick2019association}.
Moreover, mental health sequelae can prolong youth concussion patients' concussion recovery~\cite{silverberg2020management, iverson2003examination}, 
and hinder their ability to attend school and participate in normal social activities~\cite{iadevaia2015qualitative}, which can be an important way to maintain mental well-being~\cite{zamfir2015physical}.
Thus, it is important to integrate the assessment of mental health conditions into clinical follow-up procedures~\cite{gornall2021mental, russell2023incidence}. The assessment enables concussion clinicians to identify abnormalities in youth patient mental health conditions and conduct intervention, such as conducting a further assessment or referring the patient to a mental health expert. 

However, there are challenges when it comes to accessing youth concussion patients' mental health conditions. 
Most youth patients are advised to rest at home after a concussion~\cite{silverberg2023management}. 
Their mental health symptoms, such as anxiety, changes in sleep patterns, and disinhibition~\cite{brent2017psychiatric}, often manifest outside of clinics, which makes it difficult for concussion clinicians to observe these symptoms in clinical settings. 
Although clinicians can inquire about mental health symptoms during clinical visits, recall bias~\cite{frissa2016challenges,van2016accuracy} may prevent patients from providing complete and accurate information to concussion clinicians. 
Self-report questionnaires like the GAD-7 and PHQ-9 are available for anxiety and depression screening~\cite{brent2017psychiatric}.
However, clinicians often do not receive the results until the patient's next clinical visit.
Moreover, clinical follow-up visits for youth concussion patients are often insufficient and lack consistency. 
Although it is recommended to conduct follow-ups every three days after a concussion~\cite{healthmil_concussion_2024}, ~\citet{ramsay2023follow} found that actual follow-up rates vary significantly, with the lowest being only 13.2\%. 
Insufficient follow-up prevents concussion clinicians from timely collecting mental health information and assessing the youth patient’s mental health conditions. 
Accordingly, it's difficult for concussion clinicians to make timely mental health-related decisions. 
Therefore, novel approaches are needed to support clinicians in remote monitoring the mental health symptoms of youth concussion patients on time.

\subsection{AI-RPM Technologies in Clinical Settings} 
\label{sec:airpm}

 RPM includes primarily wearable devices and home-use medical equipment that collect patient health data outside of clinical settings~\cite{mendel2024advice}. These systems transmit monitored patient data to healthcare providers for real-time monitoring and assessment, thereby supporting clinicians in their decision-making~\cite{catalyst2018telehealth, baig2017systematic, us2008national}. Wearable devices, for example, serve as valuable tools for remotely monitoring patient data. Existing studies have leveraged wearable technology to track and collect key physiological data from concussion patients during their recovery period, such as physical activity levels, for recovery progress assessments~\cite{tirosh2024smartphone}. Additionally, researchers have investigated the relationship between post-concussion symptoms and physical activity in young patients. 
 For instance, \citet{yang2020bidirectional} utilized ActiGraph, a smartwatch-based device, to monitor patients’ physical activity and collect movement-related data, such as step count. 
 
 Furthermore, researchers in the Human-Computer Interaction (HCI) and Computer-Supported Cooperative Work (CSCW) fields have started exploring mHealth applications that facilitate self-management, and enhance remote communication between patients and clinicians~\cite{corwin2024using, nyapathy2019tracking, salamah2021improving, west2018common}. For instance, \citet{el2021mobile} indicated that mHealth can improve diagnostic efficiency and reduce healthcare costs for patients. RPM systems can display collected data via user interfaces for patient review~\cite{griggs2018healthcare} or integrate the data into electronic health record (EHR) systems~\cite{dinh2019wearable}. However, given the significant time constraints that clinicians already face~\cite{haikio2020expectations}, the additional data generated by RPM systems could further exacerbate these challenges.

 Recently, AI-driven tools have shown the potential to assist clinicians in reducing the cognitive load in clinical workflows by analyzing and summarizing large volumes of patient data or by presenting information in a more interpretable format ~\cite{haikio2020expectations}. 
 For instance, voice assistants powered by LLMs can interact with patients using natural language, capture subjective symptom experiences, and visualize them on a dashboard for clinicians~\cite{yang2024talk2care}. 
 Despite AI having the potential to enhance today's RPM systems, poor system design of such AI-RPM may introduce new challenges; 
 Thus, we aim to explore the best design practices to ensure these systems can enhance rather than hinder clinicians' work efficiency.

\subsection{Designing AI-based Systems for Clinical Decision Making} 
Recently, HCI researchers have had an increasing interest in the clinical-AI decision making research topic, and their works span various medical domains, such as cancer diagnosis~\cite{cai2019hello,cai2019human,denekamp2007clinical}, cardiotoxicity early detection~\cite{ahmed2024advancements}, diabetic retinopathy screening~\cite{beede2020human}, and rehabilitation assessment~\cite{lee2021human,lee2020co}.
Among some of these works, AI was leveraged for predictive analytics to help forecast patient health outcomes~\cite{zhang2024rethinking, pathak2024comparative, collins2019reporting}. \citet{dabek2022evaluation} 
developed an AI model with 88.2\% accuracy for predicting the risk of mental health sequelae within 90 days post-concussion.

Despite these advancements, the integration of AI-based decision support systems into clinical workflows remains challenging. 
One major issue is that these systems often fail to consider the needs of stakeholders during the design phase, limiting the system's  usability~\cite{cai2019hello, green2019principles, khairat2018reasons, lee2020co, romero2020lesson}. 
Moreover, clinicians remain skeptical about the accuracy of AI-generated predictions, which affects their trust in the system~\cite{bohr2020rise, yang2016investigating, elwyn2013many}.
As a result, AI-based systems may not effectively assist clinicians in making faster and more accurate decisions~\cite{antoniadi2021current}.

To address this gap, existing research has sought to incorporate human-centered design principles into AI-based system development. 
Studies have shown that involving stakeholders early in the design process and establishing continuous feedback loops~\cite{abdulaal2021clinical} can enhance the integration of AI systems into clinical workflows~\cite{wang2023human}. 
As a result, AI-based systems can be better aligned with stakeholder needs, ultimately improving their usability.
For example, Yang et al.\cite{sendak2020human} conducted multiple discussions with clinicians to understand the challenges they face during meetings, collected their feedback, and iteratively refined the design of an AI-based system based on this input. 
Similarly, Lee et al.~\cite{lee2020co} involved seven therapists throughout the design process, which led to greater satisfaction and increased adoption of the system.

These prior works inspired us to begin our project with a formative study to understand concussion clinicians' challenges and needs.
Then, we continued to involve them in the design and development process of the AI-RPM system for the mental health sequelae detection scenario.

\section{Formative Study}
\label{sec:formative}

To better understand the workflow of clinicians treating youth concussion patients with mental health issues, we conducted a formative study focusing on three main aspects: (1) how clinicians gather youth patients' mental health-related information, (2) what decisions they need to make, and (3) what the difficulties are in their current workflow. 

\subsection{Study Participants and Procedure}
\label{sec:formative-design}

\begin{table*}
  \caption{Demographics of Participants in Our Formative Study}
  \label{tab:demographics}
  \begin{tabular}{cclcc} 
    \toprule
    P\# & Gender & Department & Job Title & Year of Practice \\ 
    \midrule
    P1 & Female & Pediatric Sports Medicine & Pediatric Sports Medicine Physician & 16 years \\ 
    P2 & Female & Complex Concussion Clinic & Neuropsychologist  & 12 years \\
    P3 & Female & Concussion Clinics & Concussion Clinician & 19 years \\ 
    P4 & Male & Sports Medicine & Division Chief & 30 years \\ 
    P5 & Male & Concussion Clinics & Concussion Clinician & 16 years \\ 
    P6 & Male & Sports Medicine & Non-Operative 
Sports Medicine Doctor & 10 years \\ 
    \bottomrule
  \end{tabular}
\end{table*}

The key stakeholders in our scenarios are concussion clinicians with experience treating youth concussion patients with mental health symptoms. 
These highly specialized concussion professionals work in fast-paced environments and are often overloaded with numerous complex patient cases, which makes their time very limited. 
For this reason, we were only able to recruit six U.S.-based concussion clinicians who were available to participate in our study.
Recruitment was facilitated through professional networks within related clinical fields of domain experts in the research team via convenience sampling~\cite{sedgwick2013convenience}.
Each participant engaged in a semi-structured interview of 35-45 minutes conducted remotely via Zoom. 
Table~\ref{tab:demographics} provides details on the demographics of the participants and their clinical experience. 
This study was reviewed and approved by the first author's institution's Institutional Review Board (IRB), and the study complied with the approved procedure for ethical research practices of human subject protection.

At the beginning of each interview, we first asked the participant to recall a recent case where they encountered a youth concussion patient with mental health symptoms 
They described how they performed clinical practices, identified mental health symptoms, and executed follow-up decision-making.
While participants were sharing their experiences, we prompted participants to tell more about what type(s) of patient clinical data they collected, used, or expected to support the identification of mental health sequelae, both in and outside of clinical settings.
After that, participants shared their experience with existing tools, if any, for collecting patient data, monitoring patient mental health conditions, and identifying mental health sequelae. 
Moreover, participants shared their insights on the benefits and limitations of existing tools and the expectations of novel tools that could potentially help them.
During the interview, we asked participants to refrain from disclosing any personally identifiable information (PII). 
The complete interview protocol is provided in Appendix~\ref{sec:appendixa}.

All interviews were audio-recorded and transcribed with participants' consent. 
We employed the inductive thematic analysis approach~\cite{braun2012thematic,braun2019reflecting} to derive key themes. 
Two researchers first independently coded one session of the interview transcripts to identify meaningful segments and categorize segments into codes. 
Then, they discussed individual codes, resolved discrepancies, developed consensus codes, and applied the resulting codes to the remaining transcripts.
Finally, we iteratively grouped codes into overarching themes and refined these themes through ongoing discussions until we reached a consensus on the final set of themes.

\subsection{Findings}
\label{sec:formative-findings}

Our qualitative analysis provided a comprehensive understanding of concussion clinicians' workflow with youth concussion patients and derived three major challenges that concussion clinicians encountered during the practices: 
(1) tracking patients’ mental health-related information outside the clinic and identifying its severity;
(2) communicating effectively with patients and their family members about patients' mental health issues;
(3) assessing patients' compliance with clinical recommendations.

\subsubsection{Integrating Mental Health Screening into Concussion Management}
\label{sec:formative-findings-f1}

The majority of youth concussion patients are athletes, and concussions are commonly caused by head injuries during sports participation.
After they had a head injury, the youth patients underwent an initial assessment by primary care physicians.
According to P4, primary care physicians conducted basic evaluations and symptom scoring based on international guidelines. 
If specialized care is needed, patients are referred to specialized concussion clinicians.

The clinical tasks of concussion clinicians primarily include seeing new concussion patients and following up with existing patients, which typically follow a similar routine. 
When patients visit the clinic, concussion clinicians assess the severity of concussion symptoms with standard scales as suggested by clinical guidelines.
In addition, the clinicians conduct interview sessions with patients to collect more in-depth information about their health conditions. 
\begin{quote}
    \textit{"I guess the only way that I currently assess my patients in that way is by using the symptom score that is published in the international guidelines...... That's the symptom score that I use. And then interview. I will very commonly ask about sleep habits. I will very commonly ask about nutrition habits. I am very aggressive in making, I guess my patients go to school."} (P3)
\end{quote}
For the majority of concussion patients, the best treatment available is to have a rest both physically and cognitively. Thus, concussion clinicians will have a conversation with patients to provide personalized lifestyle recommendations and self-evaluation methodologies for concussion recovery. 
For example, P5 stated that they often provide patients with a symptom log and suggested that patients use it to track their daily activities. 
Patients will be asked to return the symptom logs during follow-up appointments, which typically occur after one week.
Symptom logs are essential for concussion clinicians to monitor their patients’ weekly health changes and concussion recovery progress.
If symptoms persist beyond 28 days, further interventions are initiated, such as referrals to neuropsychologists or specialized concussion clinics (P3, P5, P6). 
\begin{quote}
    \textit{"Yeah, after we see them (patients) for the initial visit, we generally send them home with the log and ask them to fill it out every day based on how they felt, and then bring it back when they come back for their follow up appointment so that we can track how their symptoms have been going over time."} (P5)
\end{quote}
Concussion clinicians (P1, P2, P3, P5, P6) emphasized that mental health concerns are highly prevalent among youth patients. Thus, mental health assessments are an integral part of the concussion assessment process. 
Concussion clinicians (P6, P4) typically review the patient’s mental health history, including prior diagnoses, current treatments, and medication use. 
In addition, clinicians leverage standardized questionnaires, such as the Patient Health Questionnaire-9 (PHQ-9)~\cite{levis2019accuracy} and the Generalized Anxiety Disorder 7-item Scale (GAD-7)~\cite{mossman2017generalized}, to identify potential symptoms of mental health issues and the severity of symptoms. 
Concussion clinicians (P4, P6) also observe signals of emotional distress during the conversation with patients to detect potential mental health issues and will follow up with additional mental health screening if such signals are presented.
P6 reported that there is a high portion of concussion patients develop anxiety or depression following their injury. 
In such cases, P4 often recommends cognitive behavioral therapy as the intervention strategy for the mental health issues mentioned above. 

However, if mental health symptoms are found to impede concussion recovery, concussion clinicians will need to collaborate with neurologists to determine the intervention plan, such as prescribing appropriate medications. 
For patients exhibiting severe mental health symptoms, clinicians (P2, P3) will require the patients to complete an additional suicide risk assessment. 
If no immediate suicide risk is detected, clinicians typically recommend continued rest at home with ongoing symptom monitoring. 
However, P2 mentioned that 10\% of youth concussion patients develop suicidal thoughts after a concussion. 
In such cases, clinicians refer patients to mental health specialists, such as psychiatrists or psychologists.
\begin{quote}
    \textit{``In general, of people who I see with like symptoms of like anxiety, depression, things like that after concussion. It's probably about 40 to 50 percent.''} (P6)
\end{quote}

\subsubsection{Clinicians’ Struggle to Track and Assess Mental Health Beyond the Clinic}\label{sec:trackinginfo}

As mentioned above, the first 28 days following a concussion diagnosis is a critical period of time for youth concussion patients' recovery.
However, patients only visit the concussion clinic every one or two weeks, while spending the majority of time outside the clinic, more specifically, outside of clinical supervision.
This gap creates significant challenges for concussion clinicians, who (P2, P5) emphasize the importance of monitoring a wide range of mental health symptom indicators during the concussion recovery period. 
Specifically, clinicians (P1, P2, P3, P4, P5) mentoined the importance of collecting detailed information about sleep patterns (e.g., sleep hours and wake-up times), and physical activity levels. 
P3 emphasized that the sleep patterns has not really been integrated into clinical systems.
Moreover, clinicians (P2, P3, P6) are highly interested in youth patients' academic performance (whether patients completed assignments on time) and social behavioral patterns (whether patients rest at home consistently and maintain social connections rather than withdrawing socially) during concussion recovery time at home. These academic and social behaviors provide critical insights for clinicians to assess and reflect on the severity of youth patients' mental health issues during concussion recovery.
\begin{quote}
    \textit{``From their (patients) activities, feeling [that they have] little interest or in doing activities that they normally would. Seeing difficulty or decreased performance in either activities or or school.''} (P6)
\end{quote}
Furthermore, P4 stated that to capture mental health-related anomalies, the clinician mainly relies on nonverbal cues during the clinical visits.
If any abnormalities are detected, youth patients would be asked questions directly about their mental health conditions, such as suicidal tendencies, to identify the severity of their mental health symptoms.
P5 stated that identifying the severity of the symptoms is crucial for clinicians, as it impacts the clinician's ability to adjust their treatment plans in a timely manner.
Moreover, youth concussion patients may experience impairment in functioning or panic attacks, and concussion clinicians will then focus on understanding what they had been through and what events triggered the episode. In this way clinicians can evaluate patients' mental health condition more accurately as well as design a more personalized concussion treatment plan, as suggested by P2. 
\begin{quote}
    \textit{``Anybody who has a high report of anxiety, you know, as one of the symptoms or nervousness that we ask on our concussion symptoms log or if they have an underlying diagnosis of an anxiety, disorder, depression. And then, you know, maybe a little bit more likely to refer them (patients) earlier on.''} (P5)
\end{quote}

Despite the critical need for timely mental health-related information of patients at home, concussion clinicians (P2, P5) reported significant challenges in accurately and reliably collecting the information outside the clinic.
Ideally, clinicians would benefit from having real-time access to patients’ accurate mental health-related information while patients are at home. 
However, such a system is currently unavailable.
Instead, clinicians must rely on patients’ recall of their living experiences during clinical visits to reconstruct what happened during the recovery period. 
Although these discussions could help inform clinical decision-making, patients’ memory recall is often incomplete or inaccurate, especially when clinical visits span several weeks. 
This lack of reliable data collection methodology can hinder concussion clinicians’ understanding of the patient’s recovery progress and delay timely mental health interventions, such as referrals or additional assessments.
Once youth concussion patients leave the clinic, clinicians have no way to obtain this information in a timely manner.
\begin{quote}
    \textit{``...more real-time information could be helpful. Because by the time we see them (patients) in the clinic that could be you know, weeks later they might not even remember that that (mental health sequelae-related symptoms) had happened...... It's hard for anyone to remember what they did, you know, weeks before.''} (P2)
\end{quote}

Youth concussion patients currently have some remote communication options. One way is making phone calls with the provider team and reporting patients' mental health-related information regularly to clinicians when they are outside of the clinic. 
However, clinicians (P2, P3) reported several limitations with phone communication. First, youth concussion patients rarely initiate calls to their concussion clinicians. Moreover, even if youth concussion patients make a phone call, clinicians may not always be in the office or available to answer immediately, which can delay the transmission of critical health information.

An alternative option is using MyChart, a patient portal that allows patients to access their medical records and send messages to their healthcare providers. 
Concussion clinicians (P4) strongly recommend that patients use MyChart to share health-related information, particularly when youth concussion patients are reluctant to make direct phone calls. 
Compared to phone communication, MyChart offers the advantage of asynchronous messaging, enabling clinicians to review patient information and respond at their convenience. 
Despite its potential benefits, the adoption of MyChart is limited in current clinical scenarios.
Concussion clinicians (P2, P6) noted that youth patients rarely use the messaging feature in MyChart and rarely take the initiative to contact clinicians.
Instead, parents of these youth patients are more likely to use MyChart to communicate with clinicians.

\begin{quote}
    \textit{``One in 10 [patients] maybe, or even one in 20 [patients] might send a message with a question.''} (P6)
\end{quote}

Furthermore, clinicians caution against relying on MyChart for emergencies, especially when patients experience suicidal tendencies. 
P2 warned that their busy schedules prevent them from checking messages in real time, which could lead to delays in critical interventions and put patients at risk.


\begin{quote}
    \textit{``It's not really immediate. You know, cause they (patients) might send it at night, and then we don't maybe don't see it till the next afternoon, and you know. We usually, you know, tell them if it's an emergency, don't use that method because we don't really know when we're gonna see them or be able to respond to it.''} (P2)
\end{quote}

\subsubsection{Clinicians’ Struggle to Communicate Mental Health Issues with Patients and Their Families}\label{sec:communicationissues}

Communication challenges persist not only outside the clinics but also during in-person consultations.
During in-person visits, parents often accompany their children. However, some youth concussion patients feel uncomfortable disclosing sensitive mental health issues with their parents present.
P1 described a case where a patient with severe anxiety disclosed self-harm and suicidal thoughts only after requesting to speak privately, away from their parents' presence. 


\begin{quote}
    \textit{``I stepped out of the [room], I had the mom stay in the room...... and I said, 'Is there something wrong? Is there something going on?' He said, 'Yeah, I need to talk to you privately.' ''} (P1)
\end{quote}

Clinicians suspect that a major reason for this is patients' fear of mental health stigma~\cite{apa_stigma}.
The label of having a mental illness can affect how patients are perceived socially, which could potentially isolate them from their peers.
Another reason not to disclose mental health issues is the societal expectation toward youth concussion patients.
Specifically, if youth patients are athletes who are expected to perform at a high level of competition, they may struggle with the societal expectation of their physical recovery. 

Not only do patients withhold their mental health condition, but sometimes their parents also deliberately withhold information about the patient's mental health condition as reported by concussion clinicians (P1, P4, P5).
In one case encountered by P4, the youth concussion patient was a successful athlete, and the parents downplayed their child’s mental health symptoms during clinical visits. 
These parents may worry that prolonged rest due to mental health sequelae could hinder their child’s athletic performance and career trajectory. 
So the parents downplayed their child’s mental health symptoms, hoping for their child to return to their athletic career sooner.
However, when mental health issues are not disclosed and addressed in a timely manner, the issues could get worse and lead to severe mental health issues, such as self-harm or an increased risk of suicidal behavior.
The consequences of severe mental health issues could lead to long-term physical and cognitive impairment of the patients, which might eventually hinder an athlete's ability to return to sports in their lives.
\begin{quote}
    \textit{``Their (patients) parents are very vested in the child's athletic career......the parents come in and there's obviously an agenda to minimize their symptoms. Or you know, just try not to buck the system, if it's gonna limit their (patients) activity.''} (P4)
\end{quote}

To detect undisclosed mental health issues, clinicians rely on a combination of strategies.
One approach is to build trust and collaborate directly with parents. 
Another is to observe the patient’s behavior during consultations. 
For example, clinicians (P1, P4) observe patients' behaviors, such as signs of anxiety or a persistently quiet and disengaged attitude, during clinical visits. 
Moreover, worsened concussion symptoms such as insomnia and low activity levels may indicate underlying issues beyond the concussion itself, such as mental health sequelae.
P3 stated that if a youth patient's concussion symptoms do not improve within four weeks after a concussion, it may indicate underlying mental health issues that are contributing to a prolonged recovery period.
However, observing patients' behavior or asking about their sleep patterns requires patients to be physically present during the clinical visit.
Once the patients return home, clinicians cannot access this information.
\begin{quote}
    \textit{``Because we expect 96\% of concussions to be healed by the four-week point. And so if it's not healed yet, then it's probably not the concussion, right? It's probably either a headache syndrome or a mental health concern.''} (P3)
\end{quote}

\subsubsection{Clinicians’ Struggle to Evaluate Patients’ Compliance Beyond the Clinic}\label{sec:compliance}

At the end of each clinical visit, concussion clinicians typically provide patients with recommendations related to physical activity levels or sleep schedules.
These clinical recommendations are intended to serve as the concussion treatment plan to support both physical and mental recovery after a concussion.
However, understanding whether patients adhere to these recommendations is a persistent challenge.
Although concussion clinicians (P2, P4) can rely on patients’ self-reports to assess compliance during the time period between the last visit and the current one, these reports are often unreliable. 
For instance, some patients may claim to have followed the recommendations, but their responses to mental health questionnaires reveal high levels of anxiety or depression, indicating limited or no improvement.


\begin{quote}
    \textit{“What we recommended, if they (youth patients) [said they] did it or not, they might not have done it... ... We don't know if they're following them or not. Or they come in, and they say they did, but we don't know if they did.”} (P2)
\end{quote}

However, there is currently no effective way to track patients' compliance outside the clinic. 
P4 mentioned that they often ask patients' parents about their children's adherence to the recommendations, but the information provided by the parents may also be unreliable.
One reason could be that some parents are frequently away from home due to work commitments, which leaves them with limited knowledge of patients’ daily health-related information. 
In such cases, clinicians may reach out to other family members by phone for a closer understanding of the patient’s mental health issues when they are at home. 
However, making such calls consumes a considerable amount of time in clinicians' busy schedules, and the information collected through these calls is often limited. 
\begin{quote}
    \textit{``I think sometimes the parent comes in with a child who isn't around them (parents) that much at home. Like maybe it's the working parent, and they're just sort of the driver, and they don't have a lot of information...... Sometimes I can't get good information cause they (parents) are not engaged as much as I would, you know, would hope.''} (P4)
\end{quote}

\section{Preliminary System Design}
In the formative study, we identified clinicians' urgent need to remotely monitor the mental health of youth concussion patients at home, as well as the challenges they face in clinical decision-making.
In this section, we explored the design opportunities of AI-RPM technologies in supporting clinicians’ needs and decision-making during their practice. 
Subsequently, we proposed a preliminary system design (Fig. \ref{fig:preliminary_system_design}) based on the AI-RPM technologies with detailed features. 
The goal of the design is to explore whether concussion clinicians find the AI-RPM system useful for supporting remote monitoring and decision-making, how the system can be improved, and how they would use it in practice.

\subsection{Design Considerations for AI-RPM Systems}

\subsubsection{Wearables for Tracking Patient Physiological Data}
\label{sec:wearblesforsleepandactivitiestracking}

Based on Section \ref{sec:trackinginfo}, clinicians expressed a desire to monitor patients’ at-home quantitative data, such as sleep and exercise patterns, as these factors are closely related to concussion recovery and mental health status. 
In addition, as noted in Section \ref{sec:compliance}, clinicians frequently ask patients about their clinical recommendation compliance, such as maintaining proper sleep and exercise routines.
However, clinicians currently lack effective ways to access these at-home data and often struggle to verify the accuracy of patients’ self-reported information during clinical visits.

To address these challenges, wearable devices such as smartwatches can be used to help clinicians collect real-time sleep and exercise data from patients~\cite{dias2018wearable, baig2017systematic, bate2023role}. 
Wearables offer a non-invasive and natural way to monitor patients’ health metrics without significantly disrupting their daily routines. 
Moreover, AI has the potential to process the massive amount of data collected from wearables and provide these data to clinicians for timely review.
We believe that with access to at-home patient data, clinicians can become aware of patients' health conditions in real-time, and track patients' compliance with sleep and exercise recommendations, thus ultimately having the ability to make more informed mental health-related clinical decisions, such as setting up a follow-up meeting for further assessment.

\subsubsection{LLM-based CAs for Patient Self-Reporting in Mental Health}
\label{sec:casforselfreport}

In Section \ref{sec:trackinginfo}, clinicians emphasized the need for a comprehensive understanding of patients' experiences with mental health symptoms, such as panic attacks, to provide effective care. However, patients often struggle to accurately recall these details, and such subjective experiences cannot be easily quantified using standardized scales or medical instruments. Additionally, based on Section \ref{sec:communicationissues}, concussion clinicians reported that youth concussion patients sometimes conceal their mental health issues from their parents. It was only during a private conversation with the clinician that the patient disclosed their mental health status.

Given this challenge, we hypothesize that LLM-based CAs~\cite{yang2024talk2care, mahmood2023llm, chan2024human} could help clinicians collect mental health-related information effectively. 
CAs can be deployed in patients' homes to engage in natural language conversations, ask relevant questions, and record dialogue content. 
AI can then analyze historical conversations to summarize the key information of the conversation, such as their patient's suicidal intentions. 
From the perspective of youth patients, the deployment of a CA in their rooms could increase the chance of disclosing their mental health issues, if there are any.
One study found that children perceive conversational agents as warmer and more reliable than their parents~\cite{van2023children}.
From the perspective of clinicians, gaining such key information allows them to review patients' historical conversations with CAs, gain accurate information about patients' mental health-related experiences while preserving privacy, and conduct interventions if necessary.  
Therefore, we speculate that CAs have the potential to act as a communication bridge between clinicians in the hospital and patients at home and provide patients with self-reported symptoms. Ultimately, CAs could support clinicians in conducting a more accurate assessment of patients' mental health and making timely decisions.

\subsubsection{AI-based Risk Prediction for Detecting Mental Health Sequelae}
\label{sec:aifordetection}

As we have indicated in Section \ref{sec:casforselfreport}, clinicians reported that patients and their parents sometimes deliberately conceal mental health conditions. 
In addition, clinicians noted that the severity of mental health symptoms in a patient influences their subsequent clinical decisions, such as treatment plans or providing a referral. 
However, evaluating the severity of mental health symptoms remains challenging.
Given these two challenges, we believe that AI risk prediction for disease progression could provide valuable support to clinicians in our setting. 
AI has the capability to analyze patients' health data from the EHR system and utilize algorithms to predict the likelihood of developing specific conditions over a given period~\cite{collins2019reporting}. 
In particular, recent studies have already explored AI-driven risk prediction for mental health sequelae following a concussion and have found that AI predictions can attain high accuracy~\cite{dabek2022evaluation}.
By leveraging AI risk score prediction, clinicians may no longer have to rely solely on the conversations with patients during clinical visits to make mental health assessments. 
If patients or their parents choose to withhold information, AI could help clinicians detect potential mental health sequelae in a timely manner. 
Moreover, the severity of the mental health symptoms is normally associated with the AI prediction score. 
The AI prediction score can help clinicians better assess the severity of a patient's mental health symptoms and make informed decisions accordingly.

\begin{figure*}[htbp]
  \includegraphics[draft=false,width=\textwidth]{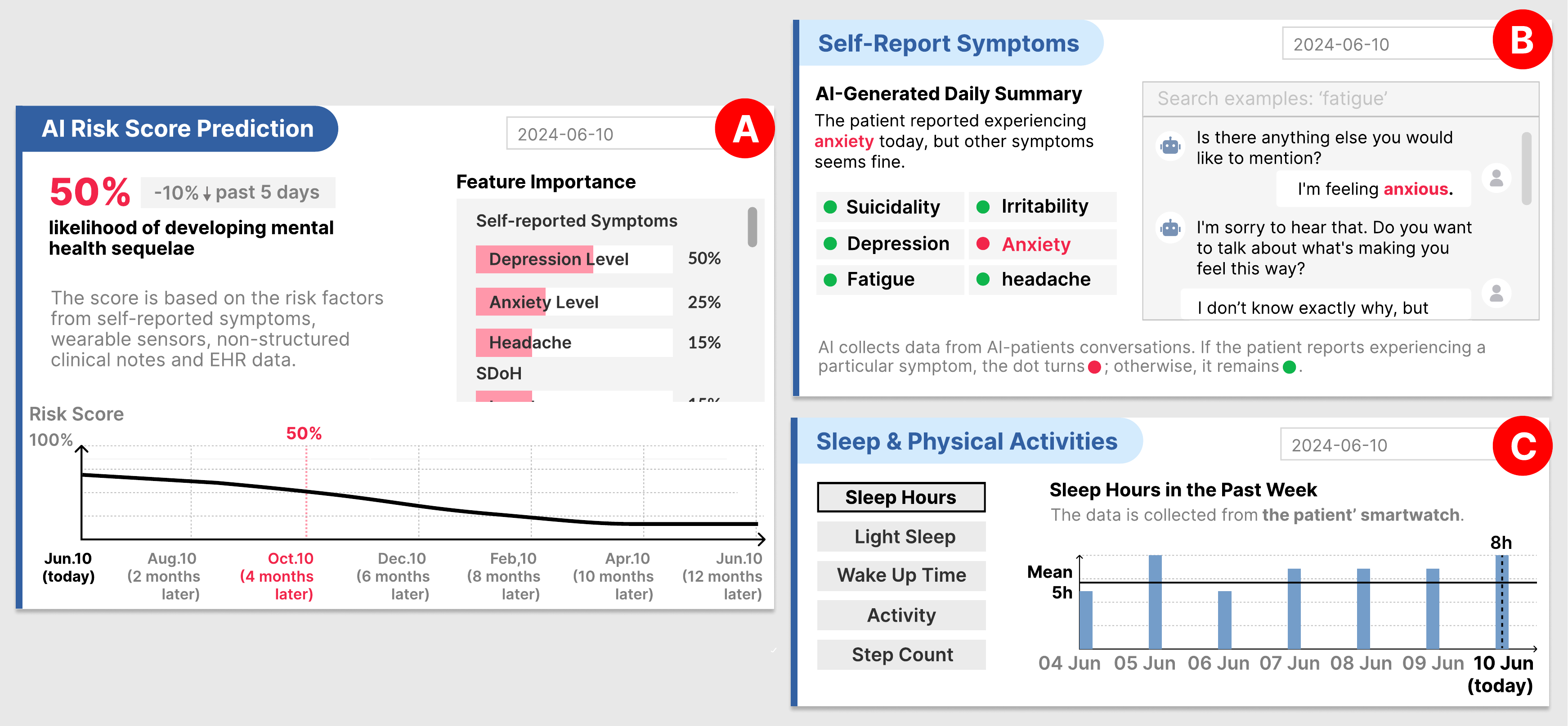}
  \caption{The preliminary system design based on concussion clinicians' challenges and needs: (A) AI risk score prediction for mental health sequelae, (B) patient self-report symptoms, (C) patient sleep and physical activities.}
  \label{fig:preliminary_system_design}
\end{figure*}

\subsection{Three Modules of Preliminary System Design Based on the Design Considerations}
Based on the potentially applicable technologies that could support concussion clinicians in addressing their challenges and supporting their decision-making, we designed three modules for our preliminary system, as shown in Fig. \ref{fig:preliminary_system_design}. 
We described each module in detail in this section.

\subsubsection{The Sleep and Physical Activities Module}
The sleep and physical activities module (Fig. \ref{fig:preliminary_system_design} B)  presents sleep (sleep hours, light sleep, and wake-up time) and activity data (step count) that clinicians want to track.
These data can be collected by wearables as we discussed on Section \ref{sec:wearblesforsleepandactivitiestracking}
Moreover, clinicians can select any category to review the patient’s relevant data from the past week. 
Clinicians can also select different dates to review patient data from other time periods. 
Additionally, the module provides the source of the data for increasing clinicians' understanding of the module and their trust towards the system. 

\subsubsection{The Self-Report Symptoms Module}
To display the patients' self-reported symptoms at home, which were collected by CAs as we proposed in Section \ref{sec:casforselfreport}, we designed the self-reported symptoms (Fig. \ref{fig:preliminary_system_design} B). 
On the right side, a history of conversations between youth concussion patients and CAs is available for concussion clinicians to review. 
Based on the conversation history, AI detects the symptoms that are mentioned by patients and highlights key information that are relevant to mental health sequelae, such as anxiety, depression, and suicidal ideation. 
If a patient self-reports symptoms within a specific category, the corresponding indicator changes from green to red.
We use the color red to catch clinicians' attention. 
To facilitate efficient navigation, clinicians can click on a category name or use the search bar to enter keywords, which prompts the right-side module to display the relevant conversation excerpts. 
Similar as the sleep and physical activities module, time selection lets clinicians view data from different periods.

\subsubsection{The AI Risk Score Prediction Module}
The AI Risk Score Prediction (Fig. \ref{fig:preliminary_system_design} A) module leverages AI's ability that we discussed in Section \ref{sec:aifordetection} to show the probability of mental health sequelae.
The prediction timeframe is within one year with a range from 0\% to 100\%, where the higher probability score signifies a higher probability of experiencing mental health sequelae for clinicians to review. 
Moreover, feature importance is used to provide data that influence the AI prediction score, which allows clinicians to decide whether to trust the AI-predicted score or not. 
To increase AI explanations and transparency, we provide text-based explanations to help clinicians understand the risk score and its source. 
At the bottom is the risk score chart. Clinicians can hover over any day on the chart to view the likelihood of developing mental health sequelae at that specific point in time. Additionally, clinicians can select different dates to review patient data from various time periods.

\section{Evaluation Study of the Preliminary System Design}

To evaluate the usefulness of our preliminary design, we conducted an evaluation study.
The goal of the evaluation study is to gather clinicians' feedback and insights on our preliminary designs with different AI-RPM technologies. Specifically, we want to explore (1) whether they find these AI-RPM technologies useful for remotely monitoring mental health-related information of youth concussion patients and (2) how to revise the preliminary system design from data type, visualization, interaction, and usability perspectives to better support clinicians in addressing challenges and decision-making in their workflow.

\subsection{Procedure}
We conducted an evaluation study with the same six clinicians from our formative study since they are the target users of the AI-RPM system. 
Due to the significant time constraints faced by concussion clinicians, each evaluation session was limited to 10 minutes with each clinician. We conducted a Zoom screen-sharing session to present our preliminary designs to each clinician and collected verbal feedback regarding data type, visualization, interaction, and usability. Meanwhile, our research team documented their responses through audio recording and note-taking. After the session, two researchers applied axial coding to identify codes and key themes. Afterward, we iterated on our preliminary designs based on our findings from this design interface evaluation. The detailed interview protocol for our evaluation study is provided in Appendix~\ref{sec:appendixb}.

\subsection{Findings} 
In this section, we present five key findings derived from concussion clinicians' feedback during the evaluation study.

\subsubsection{Improving Patient-Clinician Communication With AI Risk Score Prediction}
Concussion clinicians (P1, P5, P6) reported that if the AI risk score prediction module provides accurate data, it can be very useful for timely clinical adjustments, such as scheduling follow-up visits or providing referrals.
\begin{quote}
    \textit{``That (the AI risk score prediction) might be something helpful early on for them (concussion clinicians) to say, hey, this is somebody who probably gonna benefit from seeing another provider earlier rather than later.''} (P5)
\end{quote}
Moreover, P1 mentioned that the AI risk score prediction module is not only easy to understand but also has the potential to serve as an evidence-based tool to effectively support clinicians in communicating with patients and their parents about mental health conditions during clinical visits.
By directly presenting the risk scores of the likelihood of developing mental health sequelae, youth patients and their families may gain a clear understanding of the severity of patients' mental health conditions and be willing to talk about it, particularly for youth patients and families who are hesitant to discuss mental health issues with concussion clinicians. 
\begin{quote}
    \textit{``I think it's pretty simple, and if I was able to show that to a family... maybe something as simple as this to show [the patient's] mom... 'I'm concerned about your son... there's a 50 percent risk that [your son] may develop [mental health sequelae]. We should really be in touch with it.' Yes, I think this is great.''} (P1)
\end{quote}
However, P3 found the AI-generated risk score unhelpful because it covered one year, while she only treated youth concussion patients for four weeks.
If patients did not recover within this period, they would be referred for further treatment.

\subsubsection{Supporting Clinicians' Understanding with AI-Generated Results and Data Source Transparency}
\label{subsubsec:explanation}
Concussion clinicians expressed confusion regarding the data sources and AI results in both the AI risk score prediction modules and self-reported symptoms modules. 
When we presented the AI risk score prediction module, P6 sought for a clear explanation of the meaning behind the AI risk score despite the interface already providing key data that contribute to the risk score as well as a brief explanation of the score. 
Additionally, P2 and P4 mentioned that they were unclear about how patient's mental health symptoms were collected in the self-reported symptoms module.
Specifically, P2 was unsure whether the symptoms (e.g., suicidality and depression), were from AI risk score prediction modules or from questionnaires (e.g., GAD-7 and PHQ-9) that clinicians gave to their patients for self-assessment. However, the symptoms are collected by LLM-powered CAs through check-in conversations with youth patients.
\begin{quote}
    \textit{``So it's just my confusion. So when it says self-reported symptoms that's based on like a questionnaire that we've given them about their depression and anxiety level? Or this is what AI predicts that they have a 50 percent chance of like developing depression based on that?''} (P2)
\end{quote}

\subsubsection{Enriching Clinicians' Insights with Data from LLM-based CAs and Wearables}
\label{subsubsec:cawearables}
Clinicians reported that the data collected from CAs and wearables could be valuable in their workflow. 
Specifically, P6 mentioned that he would have a quick review of data on the self-reported symptoms module collected by CAs before each clinical visit to better understand youth patients' current health conditions. 
Moreover, concussion clinicians (P3, P6) believed that patients might be more willing to share information with a conversational agent than directly with a clinician, which makes the data collected by CAs a valuable additional data source. 
However, P3 suggested that rather than using CAs to ask patients about their mental health conditions daily, it would be more appropriate to collect the data weekly. This suggestion helps avoid constantly reminding patients of their injuries.
\begin{quote}
    \textit{``I guess from a perspective of maybe they (patients) would tell this device things that they wouldn't tell me. Possibly I could see that.''} (P6)
\end{quote}
Moreover, concussion clinicians (P3, P4, P5) particularly appreciated the data in the sleep and physical activities module collected by wearables. They highlighted that data such as sleep hours is currently missing from EHR systems, yet wearable-collected data could serve as key indicators for mental health sequelae and concussion recovery progress. 
Additionally, P3 believed that clinicians could use the sleep and activity data to assess whether patients have followed sleep and exercise recommendations, which can be useful for addressing one of their main challenges.
\begin{quote}
    \textit{``This one (the Sleep and Physical Activities panel) to me might be the most helpful [one]... I think that's always hard [to get].''} (P5)
\end{quote}

\subsubsection{Diverse Clinician Preferences in Data Details and Presentation Styles}
\label{subsubsec:detaileddata}
Concussion clinicians showed varying preferences regarding the type and level of detailed data presented within the sleep and physical activity module. 
Clinicians (P2, P3, P5) suggested incorporating napping-related data into the module, including whether the patient took naps and their duration. 
Additionally, clinicians (P1, P3) emphasized the importance of heart rate in assessing concussion recovery and monitoring mental health conditions. Specifically, they expressed interest in viewing daily resting heart rate during sleep over the course of a week. 
P1 noted that a heart rate in the fifties to sixties indicates a calm state in patients. A calm state is often associated with obtaining eight hours of quality sleep. 
In contrast, if a patient experiences fragmented or restless sleep, their resting heart rate increases. 
By analyzing physiological data, clinicians can assess if symptoms are linked to mental health or other non-concussion issues.
\begin{quote}
    \textit{``If someone's recovering, I would expect it (heart rate) to be low, [meaning] they're calm......If you (patients) are having a light, irritable kind of sleep, I would expect the resting heart rate to be in the seventies or eighties. [It means] maybe they (patients) are anxious. Maybe there's something else going on.''} (P1)
\end{quote}
Moreover, in the self-report symptoms module, clinicians (P3, P5) suggested they want to include dietary information, such as appetite or nutrition, as they noted that when youth patients experience depression or high levels of anxiety, both the patients and their parents often report decreased appetite. Importantly, instead of focusing solely on whether youth patients have mental health symptoms, concussion clinicians (P3, P5) were more concerned with the duration and severity of these symptoms. Because the severity and duration of symptoms have different impacts on clinicians' decision-making, such as treatment plans.
\begin{quote}
    \textit{``I don't think it has to be high level [of mental health symptoms]...... persistent or non-improving symptoms. That's what I look for.''} (P5)
\end{quote}
Additionally, P3 expressed a new need for a feature that could indicate how adherence to sleep and exercise recommendations impacts patients' recovery, suggesting that such insights could enhance clinical decision-making. 

\subsubsection{Concerns of Clinicians in Emergency Situations and Time Constraints}
In the self-reported symptoms module, concussion clinicians placed significant emphasis on alerts related to suicidal tendencies. P1 emphasized the need for immediate intervention when suicidal tendencies emerge in youth concussion patients. 
However, clinicians expressed several challenges. 
Firstly, there is an absence of established protocols for managing such emergency situations. 
Furthermore, P4 mentioned that the youth generation experiences frequent mood swings and may sometimes exaggerate their feelings, which could lead to false alerts.
However, failure to intervene in patients who have suicidal tendencies in a timely manner could result in a life-or-death consequence. 
To address this, clinicians (P2, P4) believed that youth patients' parents should be brought into the loop and have a partnership with parents when emergency situations arise.
\begin{quote}
    \textit{``So I think this is when you need to loop the paradigm and say, hey, your child has been identified through our software. You know, cause I think, especially since we're dealing with minors, you're gonna have to get the parents loop too.''} (P4)
\end{quote}
Moreover, clinicians (P2, P5) expressed that they have limited time to thoroughly review such detailed data due to heavy workloads.
Specifically, P1 highlighted that processing large amounts of data between tightly scheduled appointments—sometimes up to 20 patients per day—makes it nearly impossible to dedicate sufficient time and energy for in-depth analysis. 
If they have the AI-RPM system, clinicians noted that they might skim through the most relevant data quickly on the day of the appointment or shortly before to gain a basic understanding of their patient’s condition.
\begin{quote}
    \textit{``I think that's gonna that would be really  tedious to try to monitor and create extra work for us.''} (P5)
\end{quote}

\section{Refined AI-RPM System Design}
Based on the findings in the design interface evaluation session, clinicians acknowledged the value of the system with AI-RPM technologies that can potentially help them remotely monitor patients' mental health symptoms and support them to make decisions in their workflow. At the same time, they provided feedback and suggestions based on the preliminary designs. Based on their feedback, we revised our preliminary design and proposed a complete design of clinician-facing AI-RPM system design (Fig. \ref{fig:refined_system_design}).

\subsection{Reducing Cognitive Burden Through EHR Integration}
To reduce the cognitive burden placed on clinicians by the new system and ensure the system can be seamlessly integrated into clinicians' workflow, we referenced the style of current EHR systems to ensure consistency. We integrated a patient list (Fig. \ref{fig:refined_system_design} A), which already exists in the EHR system, into the left side of the interface. However, we added a red dot next to a patient's name to capture clinicians' attention if AI-RPM technologies detect abnormalities in the patient's mental health conditions. Moreover, we added a basic patient information section, which exists in the current EHR system, in the upper-left module (Fig. \ref{fig:refined_system_design} B). However, the information here is tailored to concussion clinicians' needs. These pieces of information provide clinicians with essential insights into a concussion patient's condition. In particular, the Individual Psychiatric History and Mental Health Prescription are important for concussion clinicians when detecting youth patients’ mental health sequelae and making decisions.

\subsection{The Sleep and Physical Activities Module}
Based on findings from our evaluation study, concussion clinicians affirmed the value of the data provided in the Sleep and Physical Activities module (Fig. \ref{fig:preliminary_system_design} C) in the preliminary design. However, several clinicians requested more information about data sources. In response, we designed a question mark next to the module title (Fig. \ref{fig:refined_system_design} C).
Concussion clinicians can hover over the question mark to view the data source. 
Moreover, because clinicians typically prefer a weekly overview of patient data, and some of them also want the ability to see daily metrics (e.g., heart rate), we provided the option to view data from both the last week and the previous 24 hours. 
With this option, concussion clinicians can easily track changes across different time intervals.

In addition, clinicians requested the monitoring of additional metrics, such as heart rate during sleep and nap duration. To incorporate these data points without increasing cognitive load, we organized the data into three tabs: Sleep Data, Physical Activities, and Recommendation Compliance.
The Sleep Data tab is presented by default, which allows clinicians to immediately access important metrics such as heart rate, nap duration, and sleep and wake-up times.
They can choose to view one or multiple metrics, depending on their specific needs.
Different data types are displayed in distinct colors to help clinicians visually distinguish between them more easily. 
If clinicians want to delve deeper into a patient’s sleep patterns, they can hover over a data point in the chart, then a more detailed value for that day will be displayed.

\subsection{The Self-Report Symptoms Module}
To maintain consistency with the Sleep and Physical Activities module, we added question marks indicating the data source and provided time selection options in the Self-Report Symptoms module (Fig. \ref{fig:refined_system_design} D).
Moreover, we identified emergent situations (i.e., suicidality) as both critical and challenging for clinicians to monitor, given resource constraints and the risk of false alarms. 
To address this, clinicians emphasized the importance of including parents in managing emergencies.
Thus, we introduced two sections within this module. 
On the left side, we included the CA-parent conversation history, allowing clinicians to evaluate the patient’s health conditions from family perspectives. 
This conversation history can be expanded by clicking the icon next to the title "Conversational History". 
In addition, we added two call buttons at the bottom of the module, which enables clinicians to contact either the patient or a parent to manage emergency.

Besides, clinicians confirmed the value of different categories (i.e., depression, anxiety) in this module but indicated the need for more information.
In response, we retained the original categories and introduced new ones—such as nutrition level—based on clinician feedback.
Moreover,clinicians mentioned about the importance of severity for their treatment plan.
To meet this need, we placed color-coded dots next to each symptom category (red for severe symptoms, orange for moderate, and green for none). 
To the right of each category, we provided a symptom duration indicator to show how long the reported issue has persisted, which help clinicians form a better understanding of patients health conditions.
Furthermore, clinicians can click AI Summary to view a detailed breakdown of the category, providing additional context to support their assessment.

\subsection{The Questionnaires Result Module}
In this module (Fig. \ref{fig:refined_system_design} E), the title includes an option to view data source information, which comes from patient-administered questionnaires at home. These questionnaires include GAD-7 for anxiety, PHQ-9 for depression, and PCSS for concussion severity. The middle section of the chart presents the scores for each questionnaire, along with corresponding explanations to help clinicians interpret the results. The module also displays a comparison with previous scores, providing clinicians with a sense of the patient’s recovery trend. Additionally, clinicians can click "Show Details" to view a breakdown of individual section scores within each questionnaire for a more in-depth assessment.

\subsection{The AI Risk Score Prediction Module}
The final module (Fig. \ref{fig:refined_system_design} F) presents the AI-generated prediction of a patient’s risk of developing mental health sequelae. 
Given that concussion clinicians primarily manage patient care during the first four weeks following a concussion, we adjusted the default timeframe for risk predictions from one year to four weeks, while still providing other timeframe options for selection.
Below the imeframe options, the visualization of the risk score chart remains unchanged from the preliminary design, given that clinicians reported it to be easy to interpret. 
However, clinicians expressed an interest in knowing which modules or features influence the AI-generated risk score.
To highlight the relationship between the AI risk score and other modules, we introduce Features Contributing to the AI Risk Score section at the bottom of the module.
The section includes contributing features and its corresponding module, along with a contribution rate.
Moreover, we provide visual feedback on each feature's contribution to the AI risk score. 
Both the color of the dots next to the feature and the font color of its associated percentage value dynamically change based on the degree of the contribution. 
In this way, clinicians can identify the important metrics quickly.

\section{Discussion}
 
In the following sections, we first discuss the use of AI-RPM technologies in the remote monitoring of patient mental health symptoms.
Then, we talked about collaborative design in AI-RPM systems for enhancing usability and effectiveness.
Finally, we conclude with collaborative emergency management with AI-RPM systems. 
Our work contributes to CSCW and HCI research at the intersection of health, RPM, and human-AI decision-making.

\subsection{Unraveling Patients' Intertwined Mental Health and Concussion Symptoms}

In clinical scenarios, it is not uncommon to have multiple illnesses intertwined with each other.
For instance, ~\citet{wu2024cardioai} discovered that cancer oncologists need to collaborate with cardiologists to adjust the cancer treatment plan and be constantly aware of patients' cardiotoxicity.
Nevertheless, the concussion-mental health scenario and the cancer-cardiotoxicity scenario are fundamentally different.
Concussion clinicians, according to our study, do not collaborate with mental health specialists, and neither do they diagnose mental health disorders. 
In most clinical practices, concussion clinicians assess whether a patient's mental health symptoms exceed a certain severity threshold during in-person visits. Based on this assessment, they adjust the treatment plan accordingly.

We believe that AI-RPM technologies have the potential to give clinicians insights into the severity of patients' mental health symptoms.
First, LLM-based CAs could provide insight into patients' mental health symptoms, such as the level of anxiety. 
Moreover, AI-based predictive models could help assess the likelihood of developing mental health sequelae, which can offer valuable and actionable insights into patients' mental health conditions.
As suggested by \citet{bennett2012ehrs, zhang2024rethinking}, utilizing AI predictive models to provide actionable insights can increase the usefulness of AI in clinicians' workflow. 
Future research could explore the design of AI-RPM technologies for collecting and distinguishing youth patients’ overlapping symptoms.

\subsection{Augmenting Objective Data With Subjective Experience}
Traditional RPM technologies, such as wearables, have already been used to remotely monitor concussion patients' objective data~\cite{yang2020bidirectional}.
However, collecting objective data alone is not enough in our scenarios.
As we discussed in the last section, it is critical to monitor the severity of mental health symptoms, such as anxiety and suicidal ideation.
However, these symptoms are more subjective experiences that can hardly be captured by traditional RPM technologies.
To address this gap, we propose using LLM-based CAs to gather patients' anxiety levels and other mental health conditions at home. 
LLM-based CAs could automatically detect key psychological risk factors such as suicidal thoughts from patients' conversations and generate a summary reportfor the day, which has been demoed in various prior research works in other scenarios~\cite{bartle2023machine, bartle2022second, wang2023enabling, simpson2020daisy, ma2024understanding, yang2023integrating}.

We believe that subjective information collected from LLM-powered CAs serves not only as supplementary information but also as contextual information for objective data (i.e., heart rate).
Prior HCI and CSCW research has highlighted the principle that technology design should incorporate context into data~\cite{dourish2004we, yoo2024missed}.
By combining subjective information and objective data in AI-RPM systems, clinicians could use the system to assess patients' situations more accurately and avoid unnecessary concern over what might otherwise appear to be worrisome in objective metrics alone. 

\subsection{Collaborative Design in AI-RPM Systems for Enhanced Usability and Effectiveness}

When presenting objective and subjective data in an AI-RPM system, it is important to prioritize the data presentation to the usability of the system. 
Concussion clinicians are already overwhelmed by existing EHR systems, and introducing an AI-RPM system with additional data could further increase their workload and strain limited healthcare resources.
Nevertheless, clinicians expressed a willingness to briefly review AI-RPM data before each visit to better prepare for patient consultations.
To provide information without increasing workload, designers need to collaborate with clinicians to prioritize data based on the importance of AI-RPM systems. 
Key data should be upfront while keeping secondary information hidden but accessible.
Our work offers a refined system design that presents a clear information hierarchy in a visually intuitive manner (Fig. \ref{fig:refined_system_design}), serving as a reference for future development.

Moreover, domain knowledge is essential for defining thresholds of mental health symptom severity in AI-RPM system design.
Without clear definitions of severity, AI-RPM systems may either underreport or overemphasize symptoms, leading to delayed interventions or unnecessary concerns. 
Thus, researcher and designers should closely collaborate with clinicians to establish clear thresholds for the severity of mental health symptoms so that clinicians get alert by the system only when necessary.
Such a collaborative approach among clinicians, researchers, and designers could enhance the usability and effectiveness of clinician-facing AI-RPM systems.
In the future, researchers and designers should focus on the key thresholds that influence clinician decisions.

\subsection{Collaborative Emergency Management With AI-RPM Systems}
Beyond the design of AI-RPM systems, it is also important to consider how the systems support clinicians during emergency situations in real-world practice.

In our study, concussion clinicians view AI as an important collaborative partner to support their decisions, which aligns with precious research in HCI and CSCW communities~\cite{zhang2024rethinking, hao2024advancing, yang2019unremarkable, zhang2020effect}, 
However, they expressed two main concerns in emergencies.
The first concern is the potential false alert of "suicidal ideation" in the system. 
Youth patients may often experience mood swings and express emotions with exaggeration, which could lead to false alarms.
As a result, a false alert could waste clinicians' valuable time and critical medical resources that could be allocated to other patients. 
Moreover, if false alarms occur frequently, clinicians could lose trust in the system, which eventually leads to the abandonment of such a system~\cite{liao2020questioning, amershi2019guidelines}. 
Another concern is the ambiguity of accountability that AI-RPM systems introduce. 
When an AI-RPM system triggers an emergency alert (e.g., suicidal ideation), clinicians may face time or resource constraints that delay their response. 
Such delays can lead to serious outcomes and raise questions about accountability, which may affect system adoption.


In high-risk, uncertain cases involving youth, it’s challenging to verify alerts and balance response with accountability.
We believe that addressing these two challenges requires collaboration among a broader set of stakeholders and clarifying accountability~\cite{goodman2017european} before implementing the system. 
Previous HCI research has primarily focused on involving clinicians and patients in emergency management with AI systems~\cite{wu2024cardioai, hao2024advancing}. 
We propose involving the youth patient’s family in the decision-making process alongside clinicians to manage emergencies. 
If suicidal ideation is detected, both the clinician and the patient’s family should be notified. 
Since family members are often nearby, they can respond quickly, while clinicians provide timely professional guidance.
Such collaboration helps mitigate risks from delayed responses while ensuring shared responsibility in emergencies.
Future research could explore parent-facing systems that integrate with clinicians' systems to enhance remote youth patient care.

\section{Limitations and Future Work}
Our work is not without limitations. 
First, the sample size of our participants was limited, with six concussion clinicians participating in both the formative and evaluation studies.
Recruiting these experts was particularly challenging due to their demanding workloads~\cite{wang2021brilliant,jin2020carepre} and the specialized expertise required to treat youth concussion patients with mental health issues
As a result, we recruited six highly relevant experts in this specific domain. 
Despite the limited number, our inductive thematic analysis reached thematic saturation.
Moreover, previous research employed a similar number of expert participants in related studies~\cite{cai2019hello,beede2020human,jacobs2021designing,yang2024talk2care,zhang2024rethinking}, which supports the appropriateness of our sample size. 
Future research should include a broader range of concussion clinicians across locations for greater generalizability of the findings. 

Secondly, the refined AI-RPM system has not yet been further evaluated by concussion clinicians to assess its effectiveness and usability. 
In future research, we plan to incorporate participatory design methodologies~\cite{muller1993participatory} to revise the system design, ensuring it aligns more closely with clinicians' workflows and needs. 
Moreover, our study focused on system design rather than developing an interactive prototype. The primary goal of our system design was to explore the feasibility and provide insights for further design refinements before actual development, thereby saving resources and time for both researchers and clinicians. 
Future research should investigate clinicians' pain points and needs when interacting with a functional system developed.

Finally, we want to highlight the potential of applying AI-RPM systems in varied healthcare settings. 
Several studies have explored remote monitoring in the healthcare domain. 
~\citet{wyche2024limitations} utilized mHealth to track the health conditions of type 1 diabetes among youth. 
~\citet{seals2022they} leveraged wearables to detect gait impairment, monitor patients' responses to treatment, and visualize patient data.
We believe that AI-RPM technologies have the potential to be leveraged and extended to these scenarios.
Future work should expand AI-RPM technologies to diverse healthcare settings, enabling timely interventions and improving care quality.


\section{Conclusion}

We conducted a formative study with six concussion clinicians to understand the challenges and needs during their clinical practice with youth concussion patients, particularly with respect to the patient mental health sequelae. 
Then, we derived a suite of design considerations with the use of AI-RPM technologies. Here are three design considerations. 
Firstly, we recommend the use of wearables to remotely monitor patients’ sleep and physical activity data, helping assess treatment compliance and inform mental health-related decisions. 
Secondly, we propose LLM-powered CAs to enhance mental health self-reporting, which facilitates private, natural interactions with patients and summarizes key information for clinicians. 
Thirdly, we suggest leveraging AI risk prediction in detecting concealed or worsening mental health sequelae.  Finally we delivered a clinician-facing system design as our final artifact. 
To our knowledge, this is the first study in the CSCW field to focus on the intersection of health, RPM, and human-AI decision-making, which provides a foundation for designing AI-RPM systems in varies medical scenarios.


\bibliographystyle{ACM-Reference-Format}
\bibliography{sample-base}


\begin{thebibliography}{109}


\ifx \showCODEN    \undefined \def \showCODEN     #1{\unskip}     \fi
\ifx \showDOI      \undefined \def \showDOI       #1{#1}\fi
\ifx \showISBNx    \undefined \def \showISBNx     #1{\unskip}     \fi
\ifx \showISBNxiii \undefined \def \showISBNxiii  #1{\unskip}     \fi
\ifx \showISSN     \undefined \def \showISSN      #1{\unskip}     \fi
\ifx \showLCCN     \undefined \def \showLCCN      #1{\unskip}     \fi
\ifx \shownote     \undefined \def \shownote      #1{#1}          \fi
\ifx \showarticletitle \undefined \def \showarticletitle #1{#1}   \fi
\ifx \showURL      \undefined \def \showURL       {\relax}        \fi
\providecommand\bibfield[2]{#2}
\providecommand\bibinfo[2]{#2}
\providecommand\natexlab[1]{#1}
\providecommand\showeprint[2][]{arXiv:#2}

\bibitem[Abdulaal et~al\mbox{.}(2021)]%
        {abdulaal2021clinical}
\bibfield{author}{\bibinfo{person}{Ahmed Abdulaal}, \bibinfo{person}{Aatish Patel}, \bibinfo{person}{Ahmed Al-Hindawi}, \bibinfo{person}{Esmita Charani}, \bibinfo{person}{Saleh~A Alqahtani}, \bibinfo{person}{Gary~W Davies}, \bibinfo{person}{Nabeela Mughal}, {and} \bibinfo{person}{Luke Stephen~Prockter Moore}.} \bibinfo{year}{2021}\natexlab{}.
\newblock \showarticletitle{Clinical utility and functionality of an artificial intelligence--based app to predict mortality in COVID-19: mixed methods analysis}.
\newblock \bibinfo{journal}{\emph{JMIR Formative Research}} \bibinfo{volume}{5}, \bibinfo{number}{7} (\bibinfo{year}{2021}), \bibinfo{pages}{e27992}.
\newblock


\bibitem[Agency(2024)]%
        {healthmil_concussion_2024}
\bibfield{author}{\bibinfo{person}{Defense~Health Agency}.} \bibinfo{year}{2024}\natexlab{}.
\newblock \bibinfo{booktitle}{\emph{Progressive return to activity: Primary care for acute concussion management}}.
\newblock U.S. Department of Defense.
\newblock
\urldef\tempurl%
\url{https://www.health.mil/Reference-Center/Publications/2024/02/23/Progressive-Return-to-Activity-Primary-Care-for-Acute-Concussion-Management}
\showURL{%
\tempurl}
\newblock
\shownote{Accessed: 2025-01-14}.


\bibitem[Ahmed et~al\mbox{.}(2024)]%
        {ahmed2024advancements}
\bibfield{author}{\bibinfo{person}{Bouatmane Ahmed}, \bibinfo{person}{Daaif Abdelaziz}, {and} \bibinfo{person}{Bousselham Abdelmajid}.} \bibinfo{year}{2024}\natexlab{}.
\newblock \showarticletitle{Advancements in Cardiotoxicity Detection and Assessment through Artificial Intelligence: A Comprehensive Review}. In \bibinfo{booktitle}{\emph{2024 4th International Conference on Innovative Research in Applied Science, Engineering and Technology (IRASET)}}. IEEE, \bibinfo{pages}{1--8}.
\newblock


\bibitem[Amershi et~al\mbox{.}(2019)]%
        {amershi2019guidelines}
\bibfield{author}{\bibinfo{person}{Saleema Amershi}, \bibinfo{person}{Dan Weld}, \bibinfo{person}{Mihaela Vorvoreanu}, \bibinfo{person}{Adam Fourney}, \bibinfo{person}{Besmira Nushi}, \bibinfo{person}{Penny Collisson}, \bibinfo{person}{Jina Suh}, \bibinfo{person}{Shamsi Iqbal}, \bibinfo{person}{Paul~N Bennett}, \bibinfo{person}{Kori Inkpen}, {et~al\mbox{.}}} \bibinfo{year}{2019}\natexlab{}.
\newblock \showarticletitle{Guidelines for human-AI interaction}. In \bibinfo{booktitle}{\emph{Proceedings of the 2019 chi conference on human factors in computing systems}}. \bibinfo{pages}{1--13}.
\newblock


\bibitem[Antoniadi et~al\mbox{.}(2021)]%
        {antoniadi2021current}
\bibfield{author}{\bibinfo{person}{Anna~Markella Antoniadi}, \bibinfo{person}{Yuhan Du}, \bibinfo{person}{Yasmine Guendouz}, \bibinfo{person}{Lan Wei}, \bibinfo{person}{Claudia Mazo}, \bibinfo{person}{Brett~A Becker}, {and} \bibinfo{person}{Catherine Mooney}.} \bibinfo{year}{2021}\natexlab{}.
\newblock \showarticletitle{Current challenges and future opportunities for XAI in machine learning-based clinical decision support systems: a systematic review}.
\newblock \bibinfo{journal}{\emph{Applied Sciences}} \bibinfo{volume}{11}, \bibinfo{number}{11} (\bibinfo{year}{2021}), \bibinfo{pages}{5088}.
\newblock


\bibitem[Association(2023)]%
        {apa_stigma}
\bibfield{author}{\bibinfo{person}{American~Psychiatric Association}.} \bibinfo{year}{2023}\natexlab{}.
\newblock \bibinfo{booktitle}{\emph{Stigma and Discrimination}}.
\newblock American Psychiatric Association.
\newblock
\urldef\tempurl%
\url{https://www.psychiatry.org/patients-families/stigma-and-discrimination}
\showURL{%
\tempurl}
\newblock
\shownote{Accessed: 2025-03-21}.


\bibitem[Baig et~al\mbox{.}(2017)]%
        {baig2017systematic}
\bibfield{author}{\bibinfo{person}{Mirza~Mansoor Baig}, \bibinfo{person}{Hamid GholamHosseini}, \bibinfo{person}{Aasia~A Moqeem}, \bibinfo{person}{Farhaan Mirza}, {and} \bibinfo{person}{Maria Lind{\'e}n}.} \bibinfo{year}{2017}\natexlab{}.
\newblock \showarticletitle{A systematic review of wearable patient monitoring systems--current challenges and opportunities for clinical adoption}.
\newblock \bibinfo{journal}{\emph{Journal of medical systems}}  \bibinfo{volume}{41} (\bibinfo{year}{2017}), \bibinfo{pages}{1--9}.
\newblock


\bibitem[Barnes et~al\mbox{.}(2024)]%
        {barnes2024clinician}
\bibfield{author}{\bibinfo{person}{Keely Barnes}, \bibinfo{person}{Heidi Sveistrup}, \bibinfo{person}{Mark Bayley}, \bibinfo{person}{Mary Egan}, \bibinfo{person}{Martin Bilodeau}, \bibinfo{person}{Michel Rathbone}, \bibinfo{person}{Monica Taljaard}, {and} \bibinfo{person}{Shawn Marshall}.} \bibinfo{year}{2024}\natexlab{}.
\newblock \showarticletitle{Clinician-prioritized measures to use in a remote concussion assessment: delphi study}.
\newblock \bibinfo{journal}{\emph{JMIR Formative Research}}  \bibinfo{volume}{8} (\bibinfo{year}{2024}), \bibinfo{pages}{e47246}.
\newblock


\bibitem[Bartle et~al\mbox{.}(2023)]%
        {bartle2023machine}
\bibfield{author}{\bibinfo{person}{Vince Bartle}, \bibinfo{person}{Liam Albright}, {and} \bibinfo{person}{Nicola Dell}.} \bibinfo{year}{2023}\natexlab{}.
\newblock \showarticletitle{" This machine is for the aides": tailoring voice assistant design to home health care work}. In \bibinfo{booktitle}{\emph{Proceedings of the 2023 CHI Conference on Human Factors in Computing Systems}}. \bibinfo{pages}{1--19}.
\newblock


\bibitem[Bartle et~al\mbox{.}(2022)]%
        {bartle2022second}
\bibfield{author}{\bibinfo{person}{Vince Bartle}, \bibinfo{person}{Janice Lyu}, \bibinfo{person}{Freesoul El~Shabazz-Thompson}, \bibinfo{person}{Yunmin Oh}, \bibinfo{person}{Angela~Anqi Chen}, \bibinfo{person}{Yu-Jan Chang}, \bibinfo{person}{Kenneth Holstein}, {and} \bibinfo{person}{Nicola Dell}.} \bibinfo{year}{2022}\natexlab{}.
\newblock \showarticletitle{“a second voice”: Investigating opportunities and challenges for interactive voice assistants to support home health aides}. In \bibinfo{booktitle}{\emph{Proceedings of the 2022 CHI Conference on Human Factors in Computing Systems}}. \bibinfo{pages}{1--17}.
\newblock


\bibitem[Bate et~al\mbox{.}(2023)]%
        {bate2023role}
\bibfield{author}{\bibinfo{person}{Gemma~L Bate}, \bibinfo{person}{Cameron Kirk}, \bibinfo{person}{Rana~ZU Rehman}, \bibinfo{person}{Yu Guan}, \bibinfo{person}{Alison~J Yarnall}, \bibinfo{person}{Silvia Del~Din}, {and} \bibinfo{person}{Rachael~A Lawson}.} \bibinfo{year}{2023}\natexlab{}.
\newblock \showarticletitle{The Role of Wearable Sensors to Monitor Physical Activity and Sleep Patterns in Older Adult Inpatients: A Structured Review}.
\newblock \bibinfo{journal}{\emph{Sensors}} \bibinfo{volume}{23}, \bibinfo{number}{10} (\bibinfo{year}{2023}), \bibinfo{pages}{4881}.
\newblock


\bibitem[Beede et~al\mbox{.}(2020)]%
        {beede2020human}
\bibfield{author}{\bibinfo{person}{Emma Beede}, \bibinfo{person}{Elizabeth Baylor}, \bibinfo{person}{Fred Hersch}, \bibinfo{person}{Anna Iurchenko}, \bibinfo{person}{Lauren Wilcox}, \bibinfo{person}{Paisan Ruamviboonsuk}, {and} \bibinfo{person}{Laura~M Vardoulakis}.} \bibinfo{year}{2020}\natexlab{}.
\newblock \showarticletitle{A human-centered evaluation of a deep learning system deployed in clinics for the detection of diabetic retinopathy}. In \bibinfo{booktitle}{\emph{Proceedings of the 2020 CHI conference on human factors in computing systems}}. \bibinfo{pages}{1--12}.
\newblock


\bibitem[Bennett et~al\mbox{.}(2012)]%
        {bennett2012ehrs}
\bibfield{author}{\bibinfo{person}{Casey~C Bennett}, \bibinfo{person}{Thomas~W Doub}, {and} \bibinfo{person}{Rebecca Selove}.} \bibinfo{year}{2012}\natexlab{}.
\newblock \showarticletitle{EHRs connect research and practice: Where predictive modeling, artificial intelligence, and clinical decision support intersect}.
\newblock \bibinfo{journal}{\emph{Health Policy and Technology}} \bibinfo{volume}{1}, \bibinfo{number}{2} (\bibinfo{year}{2012}), \bibinfo{pages}{105--114}.
\newblock


\bibitem[Bohr and Memarzadeh(2020)]%
        {bohr2020rise}
\bibfield{author}{\bibinfo{person}{Adam Bohr} {and} \bibinfo{person}{Kaveh Memarzadeh}.} \bibinfo{year}{2020}\natexlab{}.
\newblock \showarticletitle{The rise of artificial intelligence in healthcare applications}.
\newblock In \bibinfo{booktitle}{\emph{Artificial Intelligence in healthcare}}. \bibinfo{publisher}{Elsevier}, \bibinfo{pages}{25--60}.
\newblock


\bibitem[Braun and Clarke(2012)]%
        {braun2012thematic}
\bibfield{author}{\bibinfo{person}{Virginia Braun} {and} \bibinfo{person}{Victoria Clarke}.} \bibinfo{year}{2012}\natexlab{}.
\newblock \bibinfo{booktitle}{\emph{Thematic analysis.}}
\newblock \bibinfo{publisher}{American Psychological Association}.
\newblock


\bibitem[Braun and Clarke(2019)]%
        {braun2019reflecting}
\bibfield{author}{\bibinfo{person}{Virginia Braun} {and} \bibinfo{person}{Victoria Clarke}.} \bibinfo{year}{2019}\natexlab{}.
\newblock \showarticletitle{Reflecting on reflexive thematic analysis}.
\newblock \bibinfo{journal}{\emph{Qualitative research in sport, exercise and health}} \bibinfo{volume}{11}, \bibinfo{number}{4} (\bibinfo{year}{2019}), \bibinfo{pages}{589--597}.
\newblock


\bibitem[Brent and Max(2017)]%
        {brent2017psychiatric}
\bibfield{author}{\bibinfo{person}{David~A Brent} {and} \bibinfo{person}{Jeffrey Max}.} \bibinfo{year}{2017}\natexlab{}.
\newblock \showarticletitle{Psychiatric sequelae of concussions}.
\newblock \bibinfo{journal}{\emph{Current psychiatry reports}}  \bibinfo{volume}{19} (\bibinfo{year}{2017}), \bibinfo{pages}{1--8}.
\newblock


\bibitem[Brooks et~al\mbox{.}(2019)]%
        {brooks2019predicting}
\bibfield{author}{\bibinfo{person}{Brian~L Brooks}, \bibinfo{person}{Vickie Plourde}, \bibinfo{person}{Miriam~H Beauchamp}, \bibinfo{person}{Ken Tang}, \bibinfo{person}{Keith~Owen Yeates}, \bibinfo{person}{Michelle Keightley}, \bibinfo{person}{Peter Anderson}, \bibinfo{person}{Naddley D{\'e}sir{\'e}}, \bibinfo{person}{Nick Barrowman}, \bibinfo{person}{Roger Zemek}, {et~al\mbox{.}}} \bibinfo{year}{2019}\natexlab{}.
\newblock \showarticletitle{Predicting psychological distress after pediatric concussion}.
\newblock \bibinfo{journal}{\emph{Journal of neurotrauma}} \bibinfo{volume}{36}, \bibinfo{number}{5} (\bibinfo{year}{2019}), \bibinfo{pages}{679--685}.
\newblock


\bibitem[Bryan et~al\mbox{.}(2016)]%
        {bryan2016sports}
\bibfield{author}{\bibinfo{person}{Mersine~A Bryan}, \bibinfo{person}{Ali Rowhani-Rahbar}, \bibinfo{person}{R~Dawn Comstock}, \bibinfo{person}{Frederick Rivara}, {et~al\mbox{.}}} \bibinfo{year}{2016}\natexlab{}.
\newblock \showarticletitle{Sports-and recreation-related concussions in US youth}.
\newblock \bibinfo{journal}{\emph{Pediatrics}} \bibinfo{volume}{138}, \bibinfo{number}{1} (\bibinfo{year}{2016}).
\newblock


\bibitem[Cai et~al\mbox{.}(2019a)]%
        {cai2019human}
\bibfield{author}{\bibinfo{person}{Carrie~J Cai}, \bibinfo{person}{Emily Reif}, \bibinfo{person}{Narayan Hegde}, \bibinfo{person}{Jason Hipp}, \bibinfo{person}{Been Kim}, \bibinfo{person}{Daniel Smilkov}, \bibinfo{person}{Martin Wattenberg}, \bibinfo{person}{Fernanda Viegas}, \bibinfo{person}{Greg~S Corrado}, \bibinfo{person}{Martin~C Stumpe}, {et~al\mbox{.}}} \bibinfo{year}{2019}\natexlab{a}.
\newblock \showarticletitle{Human-centered tools for coping with imperfect algorithms during medical decision-making}. In \bibinfo{booktitle}{\emph{Proceedings of the 2019 chi conference on human factors in computing systems}}. \bibinfo{pages}{1--14}.
\newblock


\bibitem[Cai et~al\mbox{.}(2019b)]%
        {cai2019hello}
\bibfield{author}{\bibinfo{person}{Carrie~J Cai}, \bibinfo{person}{Samantha Winter}, \bibinfo{person}{David Steiner}, \bibinfo{person}{Lauren Wilcox}, {and} \bibinfo{person}{Michael Terry}.} \bibinfo{year}{2019}\natexlab{b}.
\newblock \showarticletitle{" Hello AI": uncovering the onboarding needs of medical practitioners for human-AI collaborative decision-making}.
\newblock \bibinfo{journal}{\emph{Proceedings of the ACM on Human-computer Interaction}} \bibinfo{volume}{3}, \bibinfo{number}{CSCW} (\bibinfo{year}{2019}), \bibinfo{pages}{1--24}.
\newblock


\bibitem[Carson et~al\mbox{.}(2021)]%
        {carson2021using}
\bibfield{author}{\bibinfo{person}{James~D Carson}, \bibinfo{person}{Katherine~E Healey}, {and} \bibinfo{person}{Pierre Fr{\'e}mont}.} \bibinfo{year}{2021}\natexlab{}.
\newblock \showarticletitle{Using the PHQ-9 to identify and manage depressive symptoms in patients with sport-related concussion}.
\newblock \bibinfo{journal}{\emph{Canadian family physician}} \bibinfo{volume}{67}, \bibinfo{number}{3} (\bibinfo{year}{2021}), \bibinfo{pages}{183--184}.
\newblock


\bibitem[Catalyst(2018)]%
        {catalyst2018telehealth}
\bibfield{author}{\bibinfo{person}{NEJM Catalyst}.} \bibinfo{year}{2018}\natexlab{}.
\newblock \showarticletitle{What is telehealth?}
\newblock \bibinfo{journal}{\emph{NEJM Catalyst}} \bibinfo{volume}{4}, \bibinfo{number}{1} (\bibinfo{year}{2018}).
\newblock


\bibitem[Ceniti et~al\mbox{.}(2022)]%
        {ceniti2022psychological}
\bibfield{author}{\bibinfo{person}{Amanda~K Ceniti}, \bibinfo{person}{Sakina~J Rizvi}, {and} \bibinfo{person}{Sidney~H Kennedy}.} \bibinfo{year}{2022}\natexlab{}.
\newblock \showarticletitle{Psychological and mental health sequelae of concussion: prevalence, treatment recommendations, novel biomarkers, and diagnostic challenges}.
\newblock In \bibinfo{booktitle}{\emph{Tackling the Concussion Epidemic: A Bench to Bedside Approach}}. \bibinfo{publisher}{Springer}, \bibinfo{pages}{131--151}.
\newblock


\bibitem[Chan et~al\mbox{.}(2024)]%
        {chan2024human}
\bibfield{author}{\bibinfo{person}{Szeyi Chan}, \bibinfo{person}{Shihan Fu}, \bibinfo{person}{Jiachen Li}, \bibinfo{person}{Bingsheng Yao}, \bibinfo{person}{Smit Desai}, \bibinfo{person}{Mirjana Prpa}, {and} \bibinfo{person}{Dakuo Wang}.} \bibinfo{year}{2024}\natexlab{}.
\newblock \showarticletitle{Human and llm-based voice assistant interaction: An analytical framework for user verbal and nonverbal behaviors}.
\newblock \bibinfo{journal}{\emph{arXiv preprint arXiv:2408.16465}} (\bibinfo{year}{2024}).
\newblock


\bibitem[Chanda et~al\mbox{.}(2024)]%
        {chanda2024dermatologist}
\bibfield{author}{\bibinfo{person}{Tirtha Chanda}, \bibinfo{person}{Katja Hauser}, \bibinfo{person}{Sarah Hobelsberger}, \bibinfo{person}{Tabea-Clara Bucher}, \bibinfo{person}{Carina~Nogueira Garcia}, \bibinfo{person}{Christoph Wies}, \bibinfo{person}{Harald Kittler}, \bibinfo{person}{Philipp Tschandl}, \bibinfo{person}{Cristian Navarrete-Dechent}, \bibinfo{person}{Sebastian Podlipnik}, {et~al\mbox{.}}} \bibinfo{year}{2024}\natexlab{}.
\newblock \showarticletitle{Dermatologist-like explainable AI enhances trust and confidence in diagnosing melanoma}.
\newblock \bibinfo{journal}{\emph{Nature Communications}} \bibinfo{volume}{15}, \bibinfo{number}{1} (\bibinfo{year}{2024}), \bibinfo{pages}{524}.
\newblock


\bibitem[Chung et~al\mbox{.}(2016)]%
        {chung2016boundary}
\bibfield{author}{\bibinfo{person}{Chia-Fang Chung}, \bibinfo{person}{Kristin Dew}, \bibinfo{person}{Allison Cole}, \bibinfo{person}{Jasmine Zia}, \bibinfo{person}{James Fogarty}, \bibinfo{person}{Julie~A Kientz}, {and} \bibinfo{person}{Sean~A Munson}.} \bibinfo{year}{2016}\natexlab{}.
\newblock \showarticletitle{Boundary negotiating artifacts in personal informatics: patient-provider collaboration with patient-generated data}. In \bibinfo{booktitle}{\emph{Proceedings of the 19th ACM conference on computer-supported cooperative work \& social computing}}. \bibinfo{pages}{770--786}.
\newblock


\bibitem[Collins and Moons(2019)]%
        {collins2019reporting}
\bibfield{author}{\bibinfo{person}{Gary~S Collins} {and} \bibinfo{person}{Karel~GM Moons}.} \bibinfo{year}{2019}\natexlab{}.
\newblock \showarticletitle{Reporting of artificial intelligence prediction models}.
\newblock \bibinfo{journal}{\emph{The Lancet}} \bibinfo{volume}{393}, \bibinfo{number}{10181} (\bibinfo{year}{2019}), \bibinfo{pages}{1577--1579}.
\newblock


\bibitem[Corwin et~al\mbox{.}(2024)]%
        {corwin2024using}
\bibfield{author}{\bibinfo{person}{Daniel~J Corwin}, \bibinfo{person}{Melissa Godfrey}, \bibinfo{person}{Kristy~B Arbogast}, \bibinfo{person}{Joseph~J Zorc}, \bibinfo{person}{Douglas~J Wiebe}, \bibinfo{person}{Jeremy~J Michel}, \bibinfo{person}{Ian Barnett}, \bibinfo{person}{Kelsy~M Stenger}, \bibinfo{person}{Lindsey~M Calandra}, \bibinfo{person}{Justin Cobb}, {et~al\mbox{.}}} \bibinfo{year}{2024}\natexlab{}.
\newblock \showarticletitle{Using mobile health to expedite access to specialty care for youth presenting to the emergency department with concussion at highest risk of developing persisting symptoms: a protocol paper for a non-randomised hybrid implementation-effectiveness trial}.
\newblock \bibinfo{journal}{\emph{BMJ Open}} \bibinfo{volume}{14}, \bibinfo{number}{6} (\bibinfo{year}{2024}), \bibinfo{pages}{e082644}.
\newblock


\bibitem[Dabek et~al\mbox{.}(2022)]%
        {dabek2022evaluation}
\bibfield{author}{\bibinfo{person}{Filip Dabek}, \bibinfo{person}{Peter Hoover}, \bibinfo{person}{Kendra Jorgensen-Wagers}, \bibinfo{person}{Tim Wu}, {and} \bibinfo{person}{Jesus~J Caban}.} \bibinfo{year}{2022}\natexlab{}.
\newblock \showarticletitle{Evaluation of machine learning techniques to predict the likelihood of mental health conditions following a first mTBI}.
\newblock \bibinfo{journal}{\emph{Frontiers in neurology}}  \bibinfo{volume}{12} (\bibinfo{year}{2022}), \bibinfo{pages}{769819}.
\newblock


\bibitem[Denekamp et~al\mbox{.}(2007)]%
        {denekamp2007clinical}
\bibfield{author}{\bibinfo{person}{Y Denekamp}, \bibinfo{person}{O Scott}, \bibinfo{person}{E Jacobson}, \bibinfo{person}{J Meyerovitch}, \bibinfo{person}{R Goldman}, \bibinfo{person}{H Avner-Cohen}, \bibinfo{person}{F Antebi}, {and} \bibinfo{person}{M Sherf}.} \bibinfo{year}{2007}\natexlab{}.
\newblock \showarticletitle{Clinical decision support systems for addressing information needs of physicians.}
\newblock \bibinfo{journal}{\emph{Isr Med Assoc J}} \bibinfo{volume}{9}, \bibinfo{number}{11} (\bibinfo{year}{2007}).
\newblock


\bibitem[Dias and Paulo Silva~Cunha(2018)]%
        {dias2018wearable}
\bibfield{author}{\bibinfo{person}{Duarte Dias} {and} \bibinfo{person}{Jo{\~a}o Paulo Silva~Cunha}.} \bibinfo{year}{2018}\natexlab{}.
\newblock \showarticletitle{Wearable health devices—vital sign monitoring, systems and technologies}.
\newblock \bibinfo{journal}{\emph{Sensors}} \bibinfo{volume}{18}, \bibinfo{number}{8} (\bibinfo{year}{2018}), \bibinfo{pages}{2414}.
\newblock


\bibitem[Dinh-Le et~al\mbox{.}(2019)]%
        {dinh2019wearable}
\bibfield{author}{\bibinfo{person}{Catherine Dinh-Le}, \bibinfo{person}{Rachel Chuang}, \bibinfo{person}{Sara Chokshi}, {and} \bibinfo{person}{Devin Mann}.} \bibinfo{year}{2019}\natexlab{}.
\newblock \showarticletitle{Wearable health technology and electronic health record integration: scoping review and future directions}.
\newblock \bibinfo{journal}{\emph{JMIR mHealth and uHealth}} \bibinfo{volume}{7}, \bibinfo{number}{9} (\bibinfo{year}{2019}), \bibinfo{pages}{e12861}.
\newblock


\bibitem[Donahue et~al\mbox{.}(2024)]%
        {donahue2024feasibility}
\bibfield{author}{\bibinfo{person}{Catherine~C Donahue}, \bibinfo{person}{Katherine~L Smulligan}, \bibinfo{person}{Mathew~J Wingerson}, \bibinfo{person}{Joshua~R Kniss}, \bibinfo{person}{Stacey~L Simon}, \bibinfo{person}{Julie~C Wilson}, {and} \bibinfo{person}{David~R Howell}.} \bibinfo{year}{2024}\natexlab{}.
\newblock \showarticletitle{Feasibility of at-Home Sleep Monitoring in Adolescents with and without Concussion}.
\newblock \bibinfo{journal}{\emph{Nature and Science of Sleep}} (\bibinfo{year}{2024}), \bibinfo{pages}{2257--2268}.
\newblock


\bibitem[Dourish(2004)]%
        {dourish2004we}
\bibfield{author}{\bibinfo{person}{Paul Dourish}.} \bibinfo{year}{2004}\natexlab{}.
\newblock \showarticletitle{What we talk about when we talk about context}.
\newblock \bibinfo{journal}{\emph{Personal and ubiquitous computing}}  \bibinfo{volume}{8} (\bibinfo{year}{2004}), \bibinfo{pages}{19--30}.
\newblock


\bibitem[El-Rashidy et~al\mbox{.}(2021)]%
        {el2021mobile}
\bibfield{author}{\bibinfo{person}{Nora El-Rashidy}, \bibinfo{person}{Shaker El-Sappagh}, \bibinfo{person}{SM~Riazul Islam}, \bibinfo{person}{Hazem M.~El-Bakry}, {and} \bibinfo{person}{Samir Abdelrazek}.} \bibinfo{year}{2021}\natexlab{}.
\newblock \showarticletitle{Mobile health in remote patient monitoring for chronic diseases: Principles, trends, and challenges}.
\newblock \bibinfo{journal}{\emph{Diagnostics}} \bibinfo{volume}{11}, \bibinfo{number}{4} (\bibinfo{year}{2021}), \bibinfo{pages}{607}.
\newblock


\bibitem[Elwyn et~al\mbox{.}(2013)]%
        {elwyn2013many}
\bibfield{author}{\bibinfo{person}{Glyn Elwyn}, \bibinfo{person}{Isabelle Scholl}, \bibinfo{person}{Caroline Tietbohl}, \bibinfo{person}{Mala Mann}, \bibinfo{person}{Adrian~GK Edwards}, \bibinfo{person}{Catharine Clay}, \bibinfo{person}{France L{\'e}gar{\'e}}, \bibinfo{person}{Trudy van~der Weijden}, \bibinfo{person}{Carmen~L Lewis}, \bibinfo{person}{Richard~M Wexler}, {et~al\mbox{.}}} \bibinfo{year}{2013}\natexlab{}.
\newblock \showarticletitle{“Many miles to go…”: a systematic review of the implementation of patient decision support interventions into routine clinical practice}.
\newblock \bibinfo{journal}{\emph{BMC medical informatics and decision making}}  \bibinfo{volume}{13} (\bibinfo{year}{2013}), \bibinfo{pages}{1--10}.
\newblock


\bibitem[Fogliato et~al\mbox{.}(2022)]%
        {fogliato2022goes}
\bibfield{author}{\bibinfo{person}{Riccardo Fogliato}, \bibinfo{person}{Shreya Chappidi}, \bibinfo{person}{Matthew Lungren}, \bibinfo{person}{Paul Fisher}, \bibinfo{person}{Diane Wilson}, \bibinfo{person}{Michael Fitzke}, \bibinfo{person}{Mark Parkinson}, \bibinfo{person}{Eric Horvitz}, \bibinfo{person}{Kori Inkpen}, {and} \bibinfo{person}{Besmira Nushi}.} \bibinfo{year}{2022}\natexlab{}.
\newblock \showarticletitle{Who goes first? Influences of human-AI workflow on decision making in clinical imaging}. In \bibinfo{booktitle}{\emph{Proceedings of the 2022 ACM Conference on Fairness, Accountability, and Transparency}}. \bibinfo{pages}{1362--1374}.
\newblock


\bibitem[Fralick et~al\mbox{.}(2019)]%
        {fralick2019association}
\bibfield{author}{\bibinfo{person}{Michael Fralick}, \bibinfo{person}{Eric Sy}, \bibinfo{person}{Adiba Hassan}, \bibinfo{person}{Matthew~J Burke}, \bibinfo{person}{Elizabeth Mostofsky}, {and} \bibinfo{person}{Todd Karsies}.} \bibinfo{year}{2019}\natexlab{}.
\newblock \showarticletitle{Association of concussion with the risk of suicide: a systematic review and meta-analysis}.
\newblock \bibinfo{journal}{\emph{JAMA neurology}} \bibinfo{volume}{76}, \bibinfo{number}{2} (\bibinfo{year}{2019}), \bibinfo{pages}{144--151}.
\newblock


\bibitem[Frissa et~al\mbox{.}(2016)]%
        {frissa2016challenges}
\bibfield{author}{\bibinfo{person}{Souci Frissa}, \bibinfo{person}{Stephani~L Hatch}, \bibinfo{person}{Nicola~T Fear}, \bibinfo{person}{Sarah Dorrington}, \bibinfo{person}{Laura Goodwin}, {and} \bibinfo{person}{Matthew Hotopf}.} \bibinfo{year}{2016}\natexlab{}.
\newblock \showarticletitle{Challenges in the retrospective assessment of trauma: comparing a checklist approach to a single item trauma experience screening question}.
\newblock \bibinfo{journal}{\emph{Bmc Psychiatry}}  \bibinfo{volume}{16} (\bibinfo{year}{2016}), \bibinfo{pages}{1--9}.
\newblock


\bibitem[Giebel et~al\mbox{.}(2024)]%
        {giebel2024problems}
\bibfield{author}{\bibinfo{person}{Godwin~Denk Giebel}, \bibinfo{person}{Carina Abels}, \bibinfo{person}{Felix Plescher}, \bibinfo{person}{Christian Speckemeier}, \bibinfo{person}{Nils~Frederik Schrader}, \bibinfo{person}{Kirstin B{\"o}rchers}, \bibinfo{person}{J{\"u}rgen Wasem}, \bibinfo{person}{Silke Neusser}, {and} \bibinfo{person}{Nikola Blase}.} \bibinfo{year}{2024}\natexlab{}.
\newblock \showarticletitle{Problems and Barriers Related to the Use of mHealth Apps From the Perspective of Patients: Focus Group and Interview Study}.
\newblock \bibinfo{journal}{\emph{Journal of Medical Internet Research}}  \bibinfo{volume}{26} (\bibinfo{year}{2024}), \bibinfo{pages}{e49982}.
\newblock


\bibitem[Ginsburg et~al\mbox{.}(2024)]%
        {ginsburg2024key}
\bibfield{author}{\bibinfo{person}{Geoffrey~S Ginsburg}, \bibinfo{person}{Rosalind~W Picard}, {and} \bibinfo{person}{Stephen~H Friend}.} \bibinfo{year}{2024}\natexlab{}.
\newblock \showarticletitle{Key issues as wearable digital health technologies enter clinical care}.
\newblock \bibinfo{journal}{\emph{New England Journal of Medicine}} \bibinfo{volume}{390}, \bibinfo{number}{12} (\bibinfo{year}{2024}), \bibinfo{pages}{1118--1127}.
\newblock


\bibitem[Goodman and Flaxman(2017)]%
        {goodman2017european}
\bibfield{author}{\bibinfo{person}{Bryce Goodman} {and} \bibinfo{person}{Seth Flaxman}.} \bibinfo{year}{2017}\natexlab{}.
\newblock \showarticletitle{European Union regulations on algorithmic decision-making and a “right to explanation”}.
\newblock \bibinfo{journal}{\emph{AI magazine}} \bibinfo{volume}{38}, \bibinfo{number}{3} (\bibinfo{year}{2017}), \bibinfo{pages}{50--57}.
\newblock


\bibitem[Gornall et~al\mbox{.}(2021)]%
        {gornall2021mental}
\bibfield{author}{\bibinfo{person}{Alice Gornall}, \bibinfo{person}{Michael Takagi}, \bibinfo{person}{Thilanka Morawakage}, \bibinfo{person}{Xiaomin Liu}, {and} \bibinfo{person}{Vicki Anderson}.} \bibinfo{year}{2021}\natexlab{}.
\newblock \showarticletitle{Mental health after paediatric concussion: a systematic review and meta-analysis}.
\newblock \bibinfo{journal}{\emph{British journal of sports medicine}} \bibinfo{volume}{55}, \bibinfo{number}{18} (\bibinfo{year}{2021}), \bibinfo{pages}{1048--1058}.
\newblock


\bibitem[Green and Chen(2019)]%
        {green2019principles}
\bibfield{author}{\bibinfo{person}{Ben Green} {and} \bibinfo{person}{Yiling Chen}.} \bibinfo{year}{2019}\natexlab{}.
\newblock \showarticletitle{The principles and limits of algorithm-in-the-loop decision making}.
\newblock \bibinfo{journal}{\emph{Proceedings of the ACM on Human-Computer Interaction}} \bibinfo{volume}{3}, \bibinfo{number}{CSCW} (\bibinfo{year}{2019}), \bibinfo{pages}{1--24}.
\newblock


\bibitem[Griggs et~al\mbox{.}(2018)]%
        {griggs2018healthcare}
\bibfield{author}{\bibinfo{person}{Kristen~N Griggs}, \bibinfo{person}{Olya Ossipova}, \bibinfo{person}{Christopher~P Kohlios}, \bibinfo{person}{Alessandro~N Baccarini}, \bibinfo{person}{Emily~A Howson}, {and} \bibinfo{person}{Thaier Hayajneh}.} \bibinfo{year}{2018}\natexlab{}.
\newblock \showarticletitle{Healthcare blockchain system using smart contracts for secure automated remote patient monitoring}.
\newblock \bibinfo{journal}{\emph{Journal of medical systems}}  \bibinfo{volume}{42} (\bibinfo{year}{2018}), \bibinfo{pages}{1--7}.
\newblock


\bibitem[H{\"a}iki{\"o} et~al\mbox{.}(2020)]%
        {haikio2020expectations}
\bibfield{author}{\bibinfo{person}{Juha H{\"a}iki{\"o}}, \bibinfo{person}{Sari Yli-Kauhaluoma}, \bibinfo{person}{Minna Pikkarainen}, \bibinfo{person}{Marika Iivari}, {and} \bibinfo{person}{Timo Koivum{\"a}ki}.} \bibinfo{year}{2020}\natexlab{}.
\newblock \showarticletitle{Expectations to data: Perspectives of service providers and users of future health and wellness services}.
\newblock \bibinfo{journal}{\emph{Health and Technology}}  \bibinfo{volume}{10} (\bibinfo{year}{2020}), \bibinfo{pages}{621--636}.
\newblock


\bibitem[Hao et~al\mbox{.}(2024)]%
        {hao2024advancing}
\bibfield{author}{\bibinfo{person}{Yuexing Hao}, \bibinfo{person}{Zeyu Liu}, \bibinfo{person}{Robert~N Riter}, {and} \bibinfo{person}{Saleh Kalantari}.} \bibinfo{year}{2024}\natexlab{}.
\newblock \showarticletitle{Advancing Patient-Centered Shared Decision-Making with AI Systems for Older Adult Cancer Patients}. In \bibinfo{booktitle}{\emph{Proceedings of the CHI Conference on Human Factors in Computing Systems}}. \bibinfo{pages}{1--20}.
\newblock


\bibitem[Iadevaia et~al\mbox{.}(2015)]%
        {iadevaia2015qualitative}
\bibfield{author}{\bibinfo{person}{Cheree Iadevaia}, \bibinfo{person}{Trevor Roiger}, {and} \bibinfo{person}{Mary~Beth Zwart}.} \bibinfo{year}{2015}\natexlab{}.
\newblock \showarticletitle{Qualitative examination of adolescent health-related quality of life at 1 year postconcussion}.
\newblock \bibinfo{journal}{\emph{Journal of Athletic Training}} \bibinfo{volume}{50}, \bibinfo{number}{11} (\bibinfo{year}{2015}), \bibinfo{pages}{1182--1189}.
\newblock


\bibitem[Iverson and Lange(2003)]%
        {iverson2003examination}
\bibfield{author}{\bibinfo{person}{Grant~L Iverson} {and} \bibinfo{person}{Rael~T Lange}.} \bibinfo{year}{2003}\natexlab{}.
\newblock \showarticletitle{Examination of" postconcussion-like" symptoms in a healthy sample}.
\newblock \bibinfo{journal}{\emph{Applied Neuropsychology}} \bibinfo{volume}{10}, \bibinfo{number}{3} (\bibinfo{year}{2003}), \bibinfo{pages}{137--144}.
\newblock


\bibitem[Jacobs et~al\mbox{.}(2021)]%
        {jacobs2021designing}
\bibfield{author}{\bibinfo{person}{Maia Jacobs}, \bibinfo{person}{Jeffrey He}, \bibinfo{person}{Melanie F.~Pradier}, \bibinfo{person}{Barbara Lam}, \bibinfo{person}{Andrew~C Ahn}, \bibinfo{person}{Thomas~H McCoy}, \bibinfo{person}{Roy~H Perlis}, \bibinfo{person}{Finale Doshi-Velez}, {and} \bibinfo{person}{Krzysztof~Z Gajos}.} \bibinfo{year}{2021}\natexlab{}.
\newblock \showarticletitle{Designing AI for trust and collaboration in time-constrained medical decisions: a sociotechnical lens}. In \bibinfo{booktitle}{\emph{Proceedings of the 2021 chi conference on human factors in computing systems}}. \bibinfo{pages}{1--14}.
\newblock


\bibitem[Jin et~al\mbox{.}(2020)]%
        {jin2020carepre}
\bibfield{author}{\bibinfo{person}{Zhuochen Jin}, \bibinfo{person}{Shuyuan Cui}, \bibinfo{person}{Shunan Guo}, \bibinfo{person}{David Gotz}, \bibinfo{person}{Jimeng Sun}, {and} \bibinfo{person}{Nan Cao}.} \bibinfo{year}{2020}\natexlab{}.
\newblock \showarticletitle{Carepre: An intelligent clinical decision assistance system}.
\newblock \bibinfo{journal}{\emph{ACM Transactions on Computing for Healthcare}} \bibinfo{volume}{1}, \bibinfo{number}{1} (\bibinfo{year}{2020}), \bibinfo{pages}{1--20}.
\newblock


\bibitem[Khairat et~al\mbox{.}(2018)]%
        {khairat2018reasons}
\bibfield{author}{\bibinfo{person}{Saif Khairat}, \bibinfo{person}{David Marc}, \bibinfo{person}{William Crosby}, \bibinfo{person}{Ali Al~Sanousi}, {et~al\mbox{.}}} \bibinfo{year}{2018}\natexlab{}.
\newblock \showarticletitle{Reasons for physicians not adopting clinical decision support systems: critical analysis}.
\newblock \bibinfo{journal}{\emph{JMIR medical informatics}} \bibinfo{volume}{6}, \bibinfo{number}{2} (\bibinfo{year}{2018}), \bibinfo{pages}{e8912}.
\newblock


\bibitem[Ledoux et~al\mbox{.}(2022)]%
        {ledoux2022risk}
\bibfield{author}{\bibinfo{person}{Andr{\'e}e-Anne Ledoux}, \bibinfo{person}{Richard~J Webster}, \bibinfo{person}{Anna~E Clarke}, \bibinfo{person}{Deshayne~B Fell}, \bibinfo{person}{Braden~D Knight}, \bibinfo{person}{William Gardner}, \bibinfo{person}{Paula Cloutier}, \bibinfo{person}{Clare Gray}, \bibinfo{person}{Meltem Tuna}, {and} \bibinfo{person}{Roger Zemek}.} \bibinfo{year}{2022}\natexlab{}.
\newblock \showarticletitle{Risk of mental health problems in children and youths following concussion}.
\newblock \bibinfo{journal}{\emph{JAMA network open}} \bibinfo{volume}{5}, \bibinfo{number}{3} (\bibinfo{year}{2022}), \bibinfo{pages}{e221235--e221235}.
\newblock


\bibitem[Lee et~al\mbox{.}(2020)]%
        {lee2020co}
\bibfield{author}{\bibinfo{person}{Min~Hun Lee}, \bibinfo{person}{Daniel~P Siewiorek}, \bibinfo{person}{Asim Smailagic}, \bibinfo{person}{Alexandre Bernardino}, {and} \bibinfo{person}{Sergi Berm{\'u}dez~i Badia}.} \bibinfo{year}{2020}\natexlab{}.
\newblock \showarticletitle{Co-design and evaluation of an intelligent decision support system for stroke rehabilitation assessment}.
\newblock \bibinfo{journal}{\emph{Proceedings of the ACM on Human-Computer Interaction}} \bibinfo{volume}{4}, \bibinfo{number}{CSCW2} (\bibinfo{year}{2020}), \bibinfo{pages}{1--27}.
\newblock


\bibitem[Lee et~al\mbox{.}(2021)]%
        {lee2021human}
\bibfield{author}{\bibinfo{person}{Min~Hun Lee}, \bibinfo{person}{Daniel~P Siewiorek}, \bibinfo{person}{Asim Smailagic}, \bibinfo{person}{Alexandre Bernardino}, {and} \bibinfo{person}{Sergi~Berm{\'u}dez Berm{\'u}dez~i Badia}.} \bibinfo{year}{2021}\natexlab{}.
\newblock \showarticletitle{A human-ai collaborative approach for clinical decision making on rehabilitation assessment}. In \bibinfo{booktitle}{\emph{Proceedings of the 2021 CHI conference on human factors in computing systems}}. \bibinfo{pages}{1--14}.
\newblock


\bibitem[Levis et~al\mbox{.}(2019)]%
        {levis2019accuracy}
\bibfield{author}{\bibinfo{person}{Brooke Levis}, \bibinfo{person}{Andrea Benedetti}, {and} \bibinfo{person}{Brett~D Thombs}.} \bibinfo{year}{2019}\natexlab{}.
\newblock \showarticletitle{Accuracy of Patient Health Questionnaire-9 (PHQ-9) for screening to detect major depression: individual participant data meta-analysis}.
\newblock \bibinfo{journal}{\emph{bmj}}  \bibinfo{volume}{365} (\bibinfo{year}{2019}).
\newblock


\bibitem[Liao et~al\mbox{.}(2020)]%
        {liao2020questioning}
\bibfield{author}{\bibinfo{person}{Q~Vera Liao}, \bibinfo{person}{Daniel Gruen}, {and} \bibinfo{person}{Sarah Miller}.} \bibinfo{year}{2020}\natexlab{}.
\newblock \showarticletitle{Questioning the AI: informing design practices for explainable AI user experiences}. In \bibinfo{booktitle}{\emph{Proceedings of the 2020 CHI conference on human factors in computing systems}}. \bibinfo{pages}{1--15}.
\newblock


\bibitem[Ma et~al\mbox{.}(2024)]%
        {ma2024understanding}
\bibfield{author}{\bibinfo{person}{Zilin Ma}, \bibinfo{person}{Yiyang Mei}, {and} \bibinfo{person}{Zhaoyuan Su}.} \bibinfo{year}{2024}\natexlab{}.
\newblock \showarticletitle{Understanding the benefits and challenges of using large language model-based conversational agents for mental well-being support}. In \bibinfo{booktitle}{\emph{AMIA Annual Symposium Proceedings}}, Vol.~\bibinfo{volume}{2023}. \bibinfo{pages}{1105}.
\newblock


\bibitem[Mahmood et~al\mbox{.}(2023)]%
        {mahmood2023llm}
\bibfield{author}{\bibinfo{person}{Amama Mahmood}, \bibinfo{person}{Junxiang Wang}, \bibinfo{person}{Bingsheng Yao}, \bibinfo{person}{Dakuo Wang}, {and} \bibinfo{person}{Chien-Ming Huang}.} \bibinfo{year}{2023}\natexlab{}.
\newblock \showarticletitle{LLM-Powered Conversational Voice Assistants: Interaction Patterns, Opportunities, Challenges, and Design Guidelines}.
\newblock \bibinfo{journal}{\emph{arXiv preprint arXiv:2309.13879}} (\bibinfo{year}{2023}).
\newblock


\bibitem[McLeod et~al\mbox{.}(2017)]%
        {mcleod2017rest}
\bibfield{author}{\bibinfo{person}{Tamara C~Valovich McLeod}, \bibinfo{person}{Joy~H Lewis}, \bibinfo{person}{Kate Whelihan}, {and} \bibinfo{person}{Cailee E~Welch Bacon}.} \bibinfo{year}{2017}\natexlab{}.
\newblock \showarticletitle{Rest and return to activity after sport-related concussion: a systematic review of the literature}.
\newblock \bibinfo{journal}{\emph{Journal of athletic training}} \bibinfo{volume}{52}, \bibinfo{number}{3} (\bibinfo{year}{2017}), \bibinfo{pages}{262--287}.
\newblock


\bibitem[Mendel et~al\mbox{.}(2024)]%
        {mendel2024advice}
\bibfield{author}{\bibinfo{person}{Tamir Mendel}, \bibinfo{person}{Oded Nov}, {and} \bibinfo{person}{Batia Wiesenfeld}.} \bibinfo{year}{2024}\natexlab{}.
\newblock \showarticletitle{Advice from a Doctor or AI? Understanding Willingness to Disclose Information Through Remote Patient Monitoring to Receive Health Advice}.
\newblock \bibinfo{journal}{\emph{Proceedings of the ACM on Human-Computer Interaction}} \bibinfo{volume}{8}, \bibinfo{number}{CSCW2} (\bibinfo{year}{2024}), \bibinfo{pages}{1--34}.
\newblock


\bibitem[Miller et~al\mbox{.}(2021)]%
        {miller2021association}
\bibfield{author}{\bibinfo{person}{Gabrielle~F Miller}, \bibinfo{person}{Lara DePadilla}, \bibinfo{person}{Sherry~Everett Jones}, \bibinfo{person}{Brad~N Bartholow}, \bibinfo{person}{Kelly Sarmiento}, {and} \bibinfo{person}{Matthew~J Breiding}.} \bibinfo{year}{2021}\natexlab{}.
\newblock \showarticletitle{The association between sports-or physical activity--related concussions and suicidality among US high school students, 2017}.
\newblock \bibinfo{journal}{\emph{Sports Health}} \bibinfo{volume}{13}, \bibinfo{number}{2} (\bibinfo{year}{2021}), \bibinfo{pages}{187--197}.
\newblock


\bibitem[Miller et~al\mbox{.}(2022)]%
        {miller2022salivary}
\bibfield{author}{\bibinfo{person}{Katherine~E Miller}, \bibinfo{person}{James~P MacDonald}, \bibinfo{person}{Lindsay Sullivan}, \bibinfo{person}{Lakshmi Prakruthi~Rao Venkata}, \bibinfo{person}{Junxin Shi}, \bibinfo{person}{Keith~Owen Yeates}, \bibinfo{person}{Su Chen}, \bibinfo{person}{Enas Alshaikh}, \bibinfo{person}{H~Gerry Taylor}, \bibinfo{person}{Amanda Hautmann}, {et~al\mbox{.}}} \bibinfo{year}{2022}\natexlab{}.
\newblock \showarticletitle{Salivary miRNA expression in children with persistent post-concussive symptoms}.
\newblock \bibinfo{journal}{\emph{Frontiers in public health}}  \bibinfo{volume}{10} (\bibinfo{year}{2022}), \bibinfo{pages}{890420}.
\newblock


\bibitem[Mossman et~al\mbox{.}(2017)]%
        {mossman2017generalized}
\bibfield{author}{\bibinfo{person}{Sarah~A Mossman}, \bibinfo{person}{Marissa~J Luft}, \bibinfo{person}{Heidi~K Schroeder}, \bibinfo{person}{Sara~T Varney}, \bibinfo{person}{David~E Fleck}, \bibinfo{person}{Drew~H Barzman}, \bibinfo{person}{Richard Gilman}, \bibinfo{person}{Melissa~P DelBello}, {and} \bibinfo{person}{Jeffrey~R Strawn}.} \bibinfo{year}{2017}\natexlab{}.
\newblock \showarticletitle{The Generalized Anxiety Disorder 7-item (GAD-7) scale in adolescents with generalized anxiety disorder: signal detection and validation}.
\newblock \bibinfo{journal}{\emph{Annals of clinical psychiatry: official journal of the American Academy of Clinical Psychiatrists}} \bibinfo{volume}{29}, \bibinfo{number}{4} (\bibinfo{year}{2017}), \bibinfo{pages}{227}.
\newblock


\bibitem[Muller and Kuhn(1993)]%
        {muller1993participatory}
\bibfield{author}{\bibinfo{person}{Michael~J Muller} {and} \bibinfo{person}{Sarah Kuhn}.} \bibinfo{year}{1993}\natexlab{}.
\newblock \showarticletitle{Participatory design}.
\newblock \bibinfo{journal}{\emph{Commun. ACM}} \bibinfo{volume}{36}, \bibinfo{number}{6} (\bibinfo{year}{1993}), \bibinfo{pages}{24--28}.
\newblock


\bibitem[Nyapathy and Arriaga(2019)]%
        {nyapathy2019tracking}
\bibfield{author}{\bibinfo{person}{Nikhila Nyapathy} {and} \bibinfo{person}{Rosa~I Arriaga}.} \bibinfo{year}{2019}\natexlab{}.
\newblock \showarticletitle{Tracking and reporting asthma data for children}. In \bibinfo{booktitle}{\emph{Companion Publication of the 2019 Conference on Computer Supported Cooperative Work and Social Computing}}. \bibinfo{pages}{330--334}.
\newblock


\bibitem[of~Health et~al\mbox{.}(2008)]%
        {us2008national}
\bibfield{author}{\bibinfo{person}{US~Department of Health}, \bibinfo{person}{Human Services}, {et~al\mbox{.}}} \bibinfo{year}{2008}\natexlab{}.
\newblock \bibinfo{title}{The National Alliance for Health Information Technology report to the Office of the National Coordinator for Health Information Technology on defining key health information technology terms}.
\newblock
\newblock


\bibitem[Olen and Dechert-Crooks(2017)]%
        {olen2017implantable}
\bibfield{author}{\bibinfo{person}{Melissa~M Olen} {and} \bibinfo{person}{Brynn Dechert-Crooks}.} \bibinfo{year}{2017}\natexlab{}.
\newblock \showarticletitle{Implantable cardiac devices: The utility of remote monitoring in a paediatric and CHD population}.
\newblock \bibinfo{journal}{\emph{Cardiology in the Young}} \bibinfo{volume}{27}, \bibinfo{number}{S1} (\bibinfo{year}{2017}), \bibinfo{pages}{S143--S146}.
\newblock


\bibitem[Pathak(2024)]%
        {pathak2024comparative}
\bibfield{author}{\bibinfo{person}{Rudransh Pathak}.} \bibinfo{year}{2024}\natexlab{}.
\newblock \showarticletitle{Comparative Analysis of Prominent AI Models for Enhanced Diabetes Diagnosis \& Prediction}. In \bibinfo{booktitle}{\emph{Proceedings of the 15th ACM International Conference on Bioinformatics, Computational Biology and Health Informatics}}. \bibinfo{pages}{1--1}.
\newblock


\bibitem[Purdy(2023)]%
        {purdy2023exploring}
\bibfield{author}{\bibinfo{person}{Elise Purdy}.} \bibinfo{year}{2023}\natexlab{}.
\newblock \emph{\bibinfo{title}{Exploring Screen and Social Media Use Among Young Adults With Persistent Post-Concussion Symptoms}}.
\newblock \bibinfo{thesistype}{Master's\ thesis}. \bibinfo{school}{The University of Western Ontario (Canada)}.
\newblock


\bibitem[Ramsay et~al\mbox{.}(2023)]%
        {ramsay2023follow}
\bibfield{author}{\bibinfo{person}{Scott Ramsay}, \bibinfo{person}{V~Susan Dahinten}, \bibinfo{person}{Manon Ranger}, {and} \bibinfo{person}{Shelina Babul}.} \bibinfo{year}{2023}\natexlab{}.
\newblock \showarticletitle{Follow-up visits after a concussion in the pediatric population: An integrative review}.
\newblock \bibinfo{journal}{\emph{NeuroRehabilitation}} \bibinfo{volume}{52}, \bibinfo{number}{3} (\bibinfo{year}{2023}), \bibinfo{pages}{315--328}.
\newblock


\bibitem[Rane et~al\mbox{.}(2023)]%
        {rane2023explainable}
\bibfield{author}{\bibinfo{person}{Nitin Rane}, \bibinfo{person}{Saurabh Choudhary}, {and} \bibinfo{person}{Jayesh Rane}.} \bibinfo{year}{2023}\natexlab{}.
\newblock \showarticletitle{Explainable Artificial Intelligence (XAI) in healthcare: Interpretable Models for Clinical Decision Support}.
\newblock \bibinfo{journal}{\emph{Available at SSRN 4637897}} (\bibinfo{year}{2023}).
\newblock


\bibitem[Rivara and Graham(2014)]%
        {rivara2014sports}
\bibfield{author}{\bibinfo{person}{Frederick~P Rivara} {and} \bibinfo{person}{Robert Graham}.} \bibinfo{year}{2014}\natexlab{}.
\newblock \showarticletitle{Sports-related concussions in youth: report from the Institute of Medicine and National Research Council}.
\newblock \bibinfo{journal}{\emph{Jama}} \bibinfo{volume}{311}, \bibinfo{number}{3} (\bibinfo{year}{2014}), \bibinfo{pages}{239--240}.
\newblock


\bibitem[Romero-Brufau et~al\mbox{.}(2020)]%
        {romero2020lesson}
\bibfield{author}{\bibinfo{person}{Santiago Romero-Brufau}, \bibinfo{person}{Kirk~D Wyatt}, \bibinfo{person}{Patricia Boyum}, \bibinfo{person}{Mindy Mickelson}, \bibinfo{person}{Matthew Moore}, {and} \bibinfo{person}{Cheristi Cognetta-Rieke}.} \bibinfo{year}{2020}\natexlab{}.
\newblock \showarticletitle{A lesson in implementation: a pre-post study of providers’ experience with artificial intelligence-based clinical decision support}.
\newblock \bibinfo{journal}{\emph{International journal of medical informatics}}  \bibinfo{volume}{137} (\bibinfo{year}{2020}), \bibinfo{pages}{104072}.
\newblock


\bibitem[Rose et~al\mbox{.}(2015)]%
        {rose2015diagnosis}
\bibfield{author}{\bibinfo{person}{Sean~C Rose}, \bibinfo{person}{Kevin~D Weber}, \bibinfo{person}{James~B Collen}, {and} \bibinfo{person}{Geoffrey~L Heyer}.} \bibinfo{year}{2015}\natexlab{}.
\newblock \showarticletitle{The diagnosis and management of concussion in children and adolescents}.
\newblock \bibinfo{journal}{\emph{Pediatric neurology}} \bibinfo{volume}{53}, \bibinfo{number}{2} (\bibinfo{year}{2015}), \bibinfo{pages}{108--118}.
\newblock


\bibitem[Russell et~al\mbox{.}(2023)]%
        {russell2023incidence}
\bibfield{author}{\bibinfo{person}{Kelly Russell}, \bibinfo{person}{Randy Walld}, \bibinfo{person}{James~M Bolton}, \bibinfo{person}{Daniel Chateau}, {and} \bibinfo{person}{Michael~J Ellis}.} \bibinfo{year}{2023}\natexlab{}.
\newblock \showarticletitle{Incidence of Subsequent Mental Health Disorders and Social Adversity Following Pediatric Concussion: A Longitudinal, Population-Based Study}.
\newblock \bibinfo{journal}{\emph{The Journal of Pediatrics}}  \bibinfo{volume}{259} (\bibinfo{year}{2023}), \bibinfo{pages}{113436}.
\newblock


\bibitem[Salamah et~al\mbox{.}(2021)]%
        {salamah2021improving}
\bibfield{author}{\bibinfo{person}{Yasmin Salamah}, \bibinfo{person}{Rahma~Dany Asyifa}, {and} \bibinfo{person}{Auzi Asfarian}.} \bibinfo{year}{2021}\natexlab{}.
\newblock \showarticletitle{Improving the usability of personal health record in mobile health application for people with autoimmune disease}. In \bibinfo{booktitle}{\emph{Proceedings of the Asian CHI Symposium 2021}}. \bibinfo{pages}{180--188}.
\newblock


\bibitem[Seals et~al\mbox{.}(2022)]%
        {seals2022they}
\bibfield{author}{\bibinfo{person}{Ayanna Seals}, \bibinfo{person}{Giuseppina Pilloni}, \bibinfo{person}{Jin Kim}, \bibinfo{person}{Raul Sanchez}, \bibinfo{person}{John-Ross Rizzo}, \bibinfo{person}{Leigh Charvet}, \bibinfo{person}{Oded Nov}, {and} \bibinfo{person}{Graham Dove}.} \bibinfo{year}{2022}\natexlab{}.
\newblock \showarticletitle{‘Are they doing better in the clinic or at home?’: understanding clinicians’ needs when visualizing wearable sensor data used in remote gait assessments for people with multiple sclerosis}. In \bibinfo{booktitle}{\emph{Proceedings of the 2022 CHI Conference on Human Factors in Computing Systems}}. \bibinfo{pages}{1--16}.
\newblock


\bibitem[Sedgwick(2013)]%
        {sedgwick2013convenience}
\bibfield{author}{\bibinfo{person}{Philip Sedgwick}.} \bibinfo{year}{2013}\natexlab{}.
\newblock \showarticletitle{Convenience sampling}.
\newblock \bibinfo{journal}{\emph{Bmj}}  \bibinfo{volume}{347} (\bibinfo{year}{2013}).
\newblock


\bibitem[Sendak et~al\mbox{.}(2020)]%
        {sendak2020human}
\bibfield{author}{\bibinfo{person}{Mark Sendak}, \bibinfo{person}{Madeleine~Clare Elish}, \bibinfo{person}{Michael Gao}, \bibinfo{person}{Joseph Futoma}, \bibinfo{person}{William Ratliff}, \bibinfo{person}{Marshall Nichols}, \bibinfo{person}{Armando Bedoya}, \bibinfo{person}{Suresh Balu}, {and} \bibinfo{person}{Cara O'Brien}.} \bibinfo{year}{2020}\natexlab{}.
\newblock \showarticletitle{" The human body is a black box" supporting clinical decision-making with deep learning}. In \bibinfo{booktitle}{\emph{Proceedings of the 2020 conference on fairness, accountability, and transparency}}. \bibinfo{pages}{99--109}.
\newblock


\bibitem[Shneiderman(2022)]%
        {shneiderman2022human}
\bibfield{author}{\bibinfo{person}{Ben Shneiderman}.} \bibinfo{year}{2022}\natexlab{}.
\newblock \bibinfo{booktitle}{\emph{Human-centered AI}}.
\newblock \bibinfo{publisher}{Oxford University Press}.
\newblock


\bibitem[Silverberg et~al\mbox{.}(2020)]%
        {silverberg2020management}
\bibfield{author}{\bibinfo{person}{Noah~D Silverberg}, \bibinfo{person}{Mary~Alexis Iaccarino}, \bibinfo{person}{William~J Panenka}, \bibinfo{person}{Grant~L Iverson}, \bibinfo{person}{Karen~L McCulloch}, \bibinfo{person}{Kristen Dams-O’Connor}, \bibinfo{person}{Nick Reed}, \bibinfo{person}{Michael McCrea}, \bibinfo{person}{Alison~M Cogan}, \bibinfo{person}{Min Jeong~Park Graf}, {et~al\mbox{.}}} \bibinfo{year}{2020}\natexlab{}.
\newblock \showarticletitle{Management of concussion and mild traumatic brain injury: a synthesis of practice guidelines}.
\newblock \bibinfo{journal}{\emph{Archives of Physical Medicine and Rehabilitation}} \bibinfo{volume}{101}, \bibinfo{number}{2} (\bibinfo{year}{2020}), \bibinfo{pages}{382--393}.
\newblock


\bibitem[Silverberg and Mikoli{\'c}(2023)]%
        {silverberg2023management}
\bibfield{author}{\bibinfo{person}{Noah~D Silverberg} {and} \bibinfo{person}{Ana Mikoli{\'c}}.} \bibinfo{year}{2023}\natexlab{}.
\newblock \showarticletitle{Management of psychological complications following mild traumatic brain injury}.
\newblock \bibinfo{journal}{\emph{Current Neurology and Neuroscience Reports}} \bibinfo{volume}{23}, \bibinfo{number}{3} (\bibinfo{year}{2023}), \bibinfo{pages}{49--58}.
\newblock


\bibitem[Simpson et~al\mbox{.}(2020)]%
        {simpson2020daisy}
\bibfield{author}{\bibinfo{person}{James Simpson}, \bibinfo{person}{Franziska Gaiser}, \bibinfo{person}{Miroslav Mac{\'\i}k}, {and} \bibinfo{person}{Timna Bre{\ss}gott}.} \bibinfo{year}{2020}\natexlab{}.
\newblock \showarticletitle{Daisy: a friendly conversational agent for older adults}. In \bibinfo{booktitle}{\emph{Proceedings of the 2nd conference on conversational user interfaces}}. \bibinfo{pages}{1--3}.
\newblock


\bibitem[Stein et~al\mbox{.}(2019)]%
        {stein2019risk}
\bibfield{author}{\bibinfo{person}{Murray~B Stein}, \bibinfo{person}{Sonia Jain}, \bibinfo{person}{Joseph~T Giacino}, \bibinfo{person}{Harvey Levin}, \bibinfo{person}{Sureyya Dikmen}, \bibinfo{person}{Lindsay~D Nelson}, \bibinfo{person}{Mary~J Vassar}, \bibinfo{person}{David~O Okonkwo}, \bibinfo{person}{Ramon Diaz-Arrastia}, \bibinfo{person}{Claudia~S Robertson}, {et~al\mbox{.}}} \bibinfo{year}{2019}\natexlab{}.
\newblock \showarticletitle{Risk of posttraumatic stress disorder and major depression in civilian patients after mild traumatic brain injury: a TRACK-TBI study}.
\newblock \bibinfo{journal}{\emph{JAMA psychiatry}} \bibinfo{volume}{76}, \bibinfo{number}{3} (\bibinfo{year}{2019}), \bibinfo{pages}{249--258}.
\newblock


\bibitem[Tirosh et~al\mbox{.}(2024)]%
        {tirosh2024smartphone}
\bibfield{author}{\bibinfo{person}{Oren Tirosh}, \bibinfo{person}{Jaymee Klonis}, \bibinfo{person}{Megan Hamilton}, \bibinfo{person}{John Olver}, \bibinfo{person}{Nilmini Wickramasinghe}, \bibinfo{person}{Dean Mckenzie}, \bibinfo{person}{Doa El-Ansary}, {and} \bibinfo{person}{Gavin Williams}.} \bibinfo{year}{2024}\natexlab{}.
\newblock \showarticletitle{Smartphone Technology to Facilitate Remote Postural Balance Assessment in Acute Concussion Management: Pilot Study}.
\newblock \bibinfo{journal}{\emph{Sensors}} \bibinfo{volume}{24}, \bibinfo{number}{21} (\bibinfo{year}{2024}), \bibinfo{pages}{6870}.
\newblock


\bibitem[Toresdahl et~al\mbox{.}(2021)]%
        {toresdahl2021systematic}
\bibfield{author}{\bibinfo{person}{Brett~G Toresdahl}, \bibinfo{person}{Warren~K Young}, \bibinfo{person}{Brianna Quijano}, {and} \bibinfo{person}{Daphne~A Scott}.} \bibinfo{year}{2021}\natexlab{}.
\newblock \showarticletitle{A systematic review of telehealth and sport-related concussion: baseline testing, diagnosis, and management}.
\newblock \bibinfo{journal}{\emph{HSS Journal{\textregistered}}} \bibinfo{volume}{17}, \bibinfo{number}{1} (\bibinfo{year}{2021}), \bibinfo{pages}{18--24}.
\newblock


\bibitem[Van~Brummelen et~al\mbox{.}(2023)]%
        {van2023children}
\bibfield{author}{\bibinfo{person}{Jessica Van~Brummelen}, \bibinfo{person}{Maura Kelleher}, \bibinfo{person}{Mingyan~Claire Tian}, {and} \bibinfo{person}{Nghi Nguyen}.} \bibinfo{year}{2023}\natexlab{}.
\newblock \showarticletitle{What Do Children and Parents Want and Perceive in Conversational Agents? Towards Transparent, Trustworthy, Democratized Agents}. In \bibinfo{booktitle}{\emph{Proceedings of the 22nd Annual ACM Interaction Design and Children Conference}}. \bibinfo{pages}{187--197}.
\newblock


\bibitem[Van~den Bergh and Walentynowicz(2016)]%
        {van2016accuracy}
\bibfield{author}{\bibinfo{person}{Omer Van~den Bergh} {and} \bibinfo{person}{Marta Walentynowicz}.} \bibinfo{year}{2016}\natexlab{}.
\newblock \showarticletitle{Accuracy and bias in retrospective symptom reporting}.
\newblock \bibinfo{journal}{\emph{Current opinion in psychiatry}} \bibinfo{volume}{29}, \bibinfo{number}{5} (\bibinfo{year}{2016}), \bibinfo{pages}{302--308}.
\newblock


\bibitem[van Ierssel et~al\mbox{.}(2023)]%
        {van2023clinician}
\bibfield{author}{\bibinfo{person}{Jacqueline van Ierssel}, \bibinfo{person}{Jennifer O'Neil}, \bibinfo{person}{Judy King}, \bibinfo{person}{Roger Zemek}, {and} \bibinfo{person}{Heidi Sveistrup}.} \bibinfo{year}{2023}\natexlab{}.
\newblock \showarticletitle{Clinician Perspectives on Providing Concussion Assessment and Management via Telehealth: A Mixed-Methods Study}.
\newblock \bibinfo{journal}{\emph{The Journal of Head Trauma Rehabilitation}} \bibinfo{volume}{38}, \bibinfo{number}{3} (\bibinfo{year}{2023}), \bibinfo{pages}{E233--E243}.
\newblock


\bibitem[Wang et~al\mbox{.}(2023a)]%
        {wang2023enabling}
\bibfield{author}{\bibinfo{person}{Bryan Wang}, \bibinfo{person}{Gang Li}, {and} \bibinfo{person}{Yang Li}.} \bibinfo{year}{2023}\natexlab{a}.
\newblock \showarticletitle{Enabling conversational interaction with mobile ui using large language models}. In \bibinfo{booktitle}{\emph{Proceedings of the 2023 CHI Conference on Human Factors in Computing Systems}}. \bibinfo{pages}{1--17}.
\newblock


\bibitem[Wang et~al\mbox{.}(2021)]%
        {wang2021brilliant}
\bibfield{author}{\bibinfo{person}{Dakuo Wang}, \bibinfo{person}{Liuping Wang}, \bibinfo{person}{Zhan Zhang}, \bibinfo{person}{Ding Wang}, \bibinfo{person}{Haiyi Zhu}, \bibinfo{person}{Yvonne Gao}, \bibinfo{person}{Xiangmin Fan}, {and} \bibinfo{person}{Feng Tian}.} \bibinfo{year}{2021}\natexlab{}.
\newblock \showarticletitle{“Brilliant AI doctor” in rural clinics: Challenges in AI-powered clinical decision support system deployment}. In \bibinfo{booktitle}{\emph{Proceedings of the 2021 CHI conference on human factors in computing systems}}. \bibinfo{pages}{1--18}.
\newblock


\bibitem[Wang et~al\mbox{.}(2023b)]%
        {wang2023human}
\bibfield{author}{\bibinfo{person}{Liuping Wang}, \bibinfo{person}{Zhan Zhang}, \bibinfo{person}{Dakuo Wang}, \bibinfo{person}{Weidan Cao}, \bibinfo{person}{Xiaomu Zhou}, \bibinfo{person}{Ping Zhang}, \bibinfo{person}{Jianxing Liu}, \bibinfo{person}{Xiangmin Fan}, {and} \bibinfo{person}{Feng Tian}.} \bibinfo{year}{2023}\natexlab{b}.
\newblock \showarticletitle{Human-centered design and evaluation of AI-empowered clinical decision support systems: a systematic review}.
\newblock \bibinfo{journal}{\emph{Frontiers in Computer Science}}  \bibinfo{volume}{5} (\bibinfo{year}{2023}), \bibinfo{pages}{1187299}.
\newblock


\bibitem[West et~al\mbox{.}(2018)]%
        {west2018common}
\bibfield{author}{\bibinfo{person}{Peter West}, \bibinfo{person}{Max Van~Kleek}, \bibinfo{person}{Richard Giordano}, \bibinfo{person}{Mark~J Weal}, {and} \bibinfo{person}{Nigel Shadbolt}.} \bibinfo{year}{2018}\natexlab{}.
\newblock \showarticletitle{Common barriers to the use of patient-generated data across clinical settings}. In \bibinfo{booktitle}{\emph{proceedings of the 2018 CHI Conference on Human Factors in Computing Systems}}. \bibinfo{pages}{1--13}.
\newblock


\bibitem[Wu et~al\mbox{.}(2024)]%
        {wu2024cardioai}
\bibfield{author}{\bibinfo{person}{Siyi Wu}, \bibinfo{person}{Weidan Cao}, \bibinfo{person}{Shihan Fu}, \bibinfo{person}{Bingsheng Yao}, \bibinfo{person}{Ziqi Yang}, \bibinfo{person}{Changchang Yin}, \bibinfo{person}{Varun Mishra}, \bibinfo{person}{Daniel Addison}, \bibinfo{person}{Ping Zhang}, {and} \bibinfo{person}{Dakuo Wang}.} \bibinfo{year}{2024}\natexlab{}.
\newblock \showarticletitle{CardioAI: A Multimodal AI-based System to Support Symptom Monitoring and Risk Detection of Cancer Treatment-Induced Cardiotoxicity}.
\newblock \bibinfo{journal}{\emph{arXiv preprint arXiv:2410.04592}} (\bibinfo{year}{2024}).
\newblock


\bibitem[Wyche et~al\mbox{.}(2024)]%
        {wyche2024limitations}
\bibfield{author}{\bibinfo{person}{Susan Wyche}, \bibinfo{person}{Dr~Jennifer Olson}, \bibinfo{person}{Mary Karanu}, \bibinfo{person}{Eric Omondi}, {and} \bibinfo{person}{Mike~Endovo Olonyo}.} \bibinfo{year}{2024}\natexlab{}.
\newblock \showarticletitle{Limitations of Using Mobile Phones for Managing Type 1 Diabetes (T1D) Among Youth in Low and Middle-Income Countries: Implications for mHealth}.
\newblock \bibinfo{journal}{\emph{Proceedings of the ACM on human-computer interaction}} \bibinfo{volume}{8}, \bibinfo{number}{CSCW2} (\bibinfo{year}{2024}), \bibinfo{pages}{1--19}.
\newblock


\bibitem[Yang et~al\mbox{.}(2015)]%
        {yang2015post}
\bibfield{author}{\bibinfo{person}{Jingzhen Yang}, \bibinfo{person}{Corinne Peek-Asa}, \bibinfo{person}{Tracey Covassin}, {and} \bibinfo{person}{James~C Torner}.} \bibinfo{year}{2015}\natexlab{}.
\newblock \showarticletitle{Post-concussion symptoms of depression and anxiety in division I collegiate athletes}.
\newblock \bibinfo{journal}{\emph{Developmental neuropsychology}} \bibinfo{volume}{40}, \bibinfo{number}{1} (\bibinfo{year}{2015}), \bibinfo{pages}{18--23}.
\newblock


\bibitem[Yang et~al\mbox{.}(2020)]%
        {yang2020bidirectional}
\bibfield{author}{\bibinfo{person}{Jingzhen Yang}, \bibinfo{person}{Menglin Xu}, \bibinfo{person}{Lindsay Sullivan}, \bibinfo{person}{H~Gerry Taylor}, {and} \bibinfo{person}{Keith~Owen Yeates}.} \bibinfo{year}{2020}\natexlab{}.
\newblock \showarticletitle{Bidirectional association between daily physical activity and postconcussion symptoms among youth}.
\newblock \bibinfo{journal}{\emph{JAMA Network Open}} \bibinfo{volume}{3}, \bibinfo{number}{11} (\bibinfo{year}{2020}), \bibinfo{pages}{e2027486--e2027486}.
\newblock


\bibitem[Yang et~al\mbox{.}(2021)]%
        {yang2021association}
\bibfield{author}{\bibinfo{person}{Jingzhen Yang}, \bibinfo{person}{Keith~Owen Yeates}, \bibinfo{person}{Junxin Shi}, \bibinfo{person}{Lindsay Sullivan}, \bibinfo{person}{Pengcheng Xun}, \bibinfo{person}{H~Gerry Taylor}, \bibinfo{person}{Michael Tiso}, \bibinfo{person}{Thomas Pommering}, \bibinfo{person}{James MacDonald}, \bibinfo{person}{Daniel~M Cohen}, {et~al\mbox{.}}} \bibinfo{year}{2021}\natexlab{}.
\newblock \showarticletitle{Association of self-paced physical and cognitive activities across the first week postconcussion with symptom resolution in youth}.
\newblock \bibinfo{journal}{\emph{The Journal of Head Trauma Rehabilitation}} \bibinfo{volume}{36}, \bibinfo{number}{2} (\bibinfo{year}{2021}), \bibinfo{pages}{E71--E78}.
\newblock


\bibitem[Yang et~al\mbox{.}(2019)]%
        {yang2019unremarkable}
\bibfield{author}{\bibinfo{person}{Qian Yang}, \bibinfo{person}{Aaron Steinfeld}, {and} \bibinfo{person}{John Zimmerman}.} \bibinfo{year}{2019}\natexlab{}.
\newblock \showarticletitle{Unremarkable AI: Fitting intelligent decision support into critical, clinical decision-making processes}. In \bibinfo{booktitle}{\emph{Proceedings of the 2019 CHI conference on human factors in computing systems}}. \bibinfo{pages}{1--11}.
\newblock


\bibitem[Yang et~al\mbox{.}(2016)]%
        {yang2016investigating}
\bibfield{author}{\bibinfo{person}{Qian Yang}, \bibinfo{person}{John Zimmerman}, \bibinfo{person}{Aaron Steinfeld}, \bibinfo{person}{Lisa Carey}, {and} \bibinfo{person}{James~F Antaki}.} \bibinfo{year}{2016}\natexlab{}.
\newblock \showarticletitle{Investigating the heart pump implant decision process: opportunities for decision support tools to help}. In \bibinfo{booktitle}{\emph{Proceedings of the 2016 CHI Conference on Human Factors in Computing Systems}}. \bibinfo{pages}{4477--4488}.
\newblock


\bibitem[Yang et~al\mbox{.}(2023)]%
        {yang2023integrating}
\bibfield{author}{\bibinfo{person}{Rui Yang}, \bibinfo{person}{Edison Marrese-Taylor}, \bibinfo{person}{Yuhe Ke}, \bibinfo{person}{Lechao Cheng}, \bibinfo{person}{Qingyu Chen}, {and} \bibinfo{person}{Irene Li}.} \bibinfo{year}{2023}\natexlab{}.
\newblock \showarticletitle{Integrating umls knowledge into large language models for medical question answering}.
\newblock \bibinfo{journal}{\emph{arXiv e-prints}} (\bibinfo{year}{2023}), \bibinfo{pages}{arXiv--2310}.
\newblock


\bibitem[Yang et~al\mbox{.}(2024)]%
        {yang2024talk2care}
\bibfield{author}{\bibinfo{person}{Ziqi Yang}, \bibinfo{person}{Xuhai Xu}, \bibinfo{person}{Bingsheng Yao}, \bibinfo{person}{Ethan Rogers}, \bibinfo{person}{Shao Zhang}, \bibinfo{person}{Stephen Intille}, \bibinfo{person}{Nawar Shara}, \bibinfo{person}{Guodong~Gordon Gao}, {and} \bibinfo{person}{Dakuo Wang}.} \bibinfo{year}{2024}\natexlab{}.
\newblock \showarticletitle{Talk2Care: An LLM-based Voice Assistant for Communication between Healthcare Providers and Older Adults}.
\newblock \bibinfo{journal}{\emph{Proceedings of the ACM on Interactive, Mobile, Wearable and Ubiquitous Technologies}} \bibinfo{volume}{8}, \bibinfo{number}{2} (\bibinfo{year}{2024}), \bibinfo{pages}{1--35}.
\newblock


\bibitem[Yin et~al\mbox{.}(2024)]%
        {yin2024sepsislab}
\bibfield{author}{\bibinfo{person}{Changchang Yin}, \bibinfo{person}{Pin-Yu Chen}, \bibinfo{person}{Bingsheng Yao}, \bibinfo{person}{Dakuo Wang}, \bibinfo{person}{Jeffrey Caterino}, {and} \bibinfo{person}{Ping Zhang}.} \bibinfo{year}{2024}\natexlab{}.
\newblock \showarticletitle{Sepsislab: Early sepsis prediction with uncertainty quantification and active sensing}. In \bibinfo{booktitle}{\emph{Proceedings of the 30th ACM SIGKDD Conference on Knowledge Discovery and Data Mining}}. \bibinfo{pages}{6158--6168}.
\newblock


\bibitem[Yoo et~al\mbox{.}(2024)]%
        {yoo2024missed}
\bibfield{author}{\bibinfo{person}{Dong~Whi Yoo}, \bibinfo{person}{Hayoung Woo}, \bibinfo{person}{Sachin~R Pendse}, \bibinfo{person}{Nathaniel~Young Lu}, \bibinfo{person}{Michael~L Birnbaum}, \bibinfo{person}{Gregory~D Abowd}, {and} \bibinfo{person}{Munmun De~Choudhury}.} \bibinfo{year}{2024}\natexlab{}.
\newblock \showarticletitle{Missed opportunities for human-centered AI research: Understanding stakeholder collaboration in mental health AI research}.
\newblock \bibinfo{journal}{\emph{Proceedings of the ACM on Human-Computer Interaction}} \bibinfo{volume}{8}, \bibinfo{number}{CSCW1} (\bibinfo{year}{2024}), \bibinfo{pages}{1--24}.
\newblock


\bibitem[Zamfir(2015)]%
        {zamfir2015physical}
\bibfield{author}{\bibinfo{person}{MV Zamfir}.} \bibinfo{year}{2015}\natexlab{}.
\newblock \showarticletitle{The Physical and Mental Benefits of Socialization}.
\newblock \bibinfo{journal}{\emph{Marathon-Revista științelor motricit{\u{a}}ții umane}} \bibinfo{volume}{12}, \bibinfo{number}{1} (\bibinfo{year}{2015}).
\newblock


\bibitem[Zhang et~al\mbox{.}(2024)]%
        {zhang2024rethinking}
\bibfield{author}{\bibinfo{person}{Shao Zhang}, \bibinfo{person}{Jianing Yu}, \bibinfo{person}{Xuhai Xu}, \bibinfo{person}{Changchang Yin}, \bibinfo{person}{Yuxuan Lu}, \bibinfo{person}{Bingsheng Yao}, \bibinfo{person}{Melanie Tory}, \bibinfo{person}{Lace~M Padilla}, \bibinfo{person}{Jeffrey Caterino}, \bibinfo{person}{Ping Zhang}, {et~al\mbox{.}}} \bibinfo{year}{2024}\natexlab{}.
\newblock \showarticletitle{Rethinking human-ai collaboration in complex medical decision making: A case study in sepsis diagnosis}. In \bibinfo{booktitle}{\emph{Proceedings of the CHI Conference on Human Factors in Computing Systems}}. \bibinfo{pages}{1--18}.
\newblock


\bibitem[Zhang et~al\mbox{.}(2020)]%
        {zhang2020effect}
\bibfield{author}{\bibinfo{person}{Yunfeng Zhang}, \bibinfo{person}{Q~Vera Liao}, {and} \bibinfo{person}{Rachel~KE Bellamy}.} \bibinfo{year}{2020}\natexlab{}.
\newblock \showarticletitle{Effect of confidence and explanation on accuracy and trust calibration in AI-assisted decision making}. In \bibinfo{booktitle}{\emph{Proceedings of the 2020 conference on fairness, accountability, and transparency}}. \bibinfo{pages}{295--305}.
\newblock


\end{thebibliography}
\appendix
\section{APPENDIX: INTERVIEW SCRIPTS FOR FORMATIVE STUDY}
\label{sec:appendixa}
We begin with a brief introduction of our research and study procedure.

\textbf{Background}

(1) Could you please tell me your years of practice, department, and specialization? 

(2) How many patients do you see each year? How many of them have mental health symptoms? How do you handle them (maybe they don’t need to refer)?

\textbf{A Recent Experience}

(1) Can you think of a recent time when you had a young concussion patient (11-17 years old) and thought they might have mental health issues? How did you find out?

\textbf{Information Collection and Monitoring}

(1) (If not mentioned in the previous question) For that patient in your story, before the final referral, what symptom-related information makes you think it might be a mental health issue? How did you deal with it?

(2) (If they mention any at-home symptoms) What kind of symptom-related information do you find helpful when patients are resting at home (smartwatch and smart speaker)? (If not mentioned) Do you need any at-home symptoms? (e.g.,  patients' mental status or sleep patterns at home ) 

(3) (If not mentioned) How do their parents get involved in this process? 

\textbf{Challenges and Workaround Strategies}

(1) What challenges do you have when it comes to collecting symptom-related information from patients, managing that information, or making mental health decisions (e.g., further mental health testing or referring to a mental health expert)?

\textbf{Technology (Expectations and Concerns)}

(1) Do you use any existing tools or features in EHR (e.g., EPIC) in the youth patients' mental health decision-making? If so, what are they, and how do you like them?

\section{APPENDIX: INTERVIEW SCRIPTS FOR EVALUATION STUDY}
\label{sec:appendixb}

(1) This section (Fig.1) shows the probability of mental health sequelae within the next 15 days with a range from 0\% to 100\%, where the higher probability score signifies a higher probability of experiencing mental health sequelae. Feature importance scores show the contributing factors to the probability. The risk score chart below shows the risk scores for the past ten days, today, and the next 15 days. We’d love your comments. What do you think about it? What other information or visualization would you like to see?

(2) (Less familiar, need extra emphasis) In this section, we will use a smart speaker to have conversations with patients. The conversation history is shown on the right. On the left is the symptom overview section. If patient-reported symptoms toward a particular category, the dot for that category will change from green to red as a highlight for your review. You can click on the category name or use the search bar to type in keywords,  then the corresponding conversation content will be shown on the right. What do you think about it? What other information or visualization would you like to see?

(3) (More familiar, need less time) The data in this section is collected from patients' smartwatches. You can click on other data buttons to view corresponding recorded data in detail. What do you think about it? What other information or visualization would you like to see?

(4) Above these three sections, what other information would you need to help make mental health decisions?

\textbf{Closing Question}
(1) Is there anything else you’d like to share or a final question to ask us? Let us know or reach out via follow-up email.

\end{document}